\documentclass[12pt]{article}   

\usepackage{color}
\usepackage{float}
\usepackage{hyperref}
\hypersetup{
    bookmarks=true,         
    unicode=false,          
    pdftoolbar=true,        
    pdfmenubar=true,        
    pdffitwindow=false,     
    pdfstartview={FitH},    
    pdftitle={},    
    pdfauthor={},     
    pdfsubject={},   
    pdfcreator={},   
    pdfproducer={}, 
    pdfkeywords={key1} {key2} {key3}, 
    pdfnewwindow=true,      
    colorlinks=true,       
    linkcolor=blue,          
    citecolor=blue,        
    filecolor=blue,      
    urlcolor=blue           
}
\usepackage[normalem]{ulem}
\usepackage{cite}
\usepackage{lscape}
\usepackage{amsfonts}
\usepackage{amssymb}
\usepackage{amsbsy} 
\usepackage{amsmath}
\usepackage{multirow}

\usepackage{mathptmx}       
\usepackage{helvet}         
\usepackage{courier}        
\usepackage{type1cm}        

\usepackage{makeidx}         
\usepackage{graphicx}        
\usepackage{multicol}        
\usepackage[bottom]{footmisc}

\makeindex             

\usepackage{url}
\usepackage{array}
\usepackage[T1]{fontenc}
\usepackage{enumitem}
\usepackage{authblk} 
\usepackage{natbib}
\usepackage{lineno} 

\newcommand{\IG}{\includegraphics}

\newcommand{\remove}[1]{}

\begin{document}

\title{Quantifying vegetation biophysical variables from imaging spectroscopy data: a review on retrieval methods}

\author{Verrelst, Jochem;  Malenovsk\'y, Zbyn\v{e}k; Van der Tol, Christiaan; Camps-Valls, Gustau; Gastellu-Etchegorry, Jean-Philippe; Lewis, Philip; North, Peter; Moreno, Jose}
\affil[1]{Image Processing Laboratory, Universitat de Val{\`{e}}ncia, Spain.}
\affil[2]{Surveying and Spatial Sciences Group, School of Technology, Environments and Design, University of Tasmania, Private Bag 76, TAS 7001 Hobart, Australia}
\affil[3]{Global Change Research Institute CAS, Remote Sensing Department, B\v{e}lidla 986/4a, 603 00 Brno, Czech Republic}
\affil[4]{USRA/GESTAR, NASA Goddard Space Flight Center, Biospheric Sciences Laboratory, 8800 Greenbelt Rd, Greenbelt, MD 20771, USA}
\affil[5]{Department of Water Resources, Faculty ITC, University of Twente, P.O. Box 217, 7500 AE Enschede, The Netherlands}
\affil[6]{Centre d'Etudes Spatiales de la Biosph\`{e}re - UPS, CNES, CNRS, IRD,
Universit\'{e} de Toulouse, 31401 Toulouse Cedex 9, France}
\affil[7]{Department of Geography, University College London, Pearson Building, Gower Street, London WC1E 6BT, UK}
\affil[8]{Department of Geography, Swansea University, SA2 8PP Swansea, UK}

\date{{\bf Preprint. Manuscript published in Surv Geophys 40, 589--629 (2019). https://doi.org/10.1007/s10712-018-9478-y}}

\setcounter{Maxaffil}{0}
\renewcommand\Affilfont{\itshape\scriptsize}

\maketitle

\begin{abstract}
An unprecedented spectroscopic data stream will soon become available with forthcoming Earth-observing satellite missions equipped with imaging spectroradiometers. This data stream will open up a vast array of opportunities to quantify a diversity of biochemical and structural vegetation properties. The processing requirements for such large data streams require reliable retrieval techniques enabling the spatio-temporally explicit quantification of biophysical variables. With the aim of preparing for this new era of Earth observation, this review summarizes the state-of-the-art retrieval methods that have been applied in experimental imaging spectroscopy studies inferring all kinds of vegetation biophysical variables. Identified retrieval methods are categorized into: (1) parametric regression, including vegetation indices, shape indices and spectral transformations; (2) non-parametric regression, including linear and non-linear machine learning regression algorithms; (3) physically-based, including inversion of radiative transfer models (RTMs) using numerical optimization and look-up table approaches; and (4) hybrid regression methods, which combine RTM simulations with machine learning regression methods. For each of these categories, an overview of widely applied methods with application to mapping vegetation properties is given. In view of processing imaging spectroscopy data, a critical aspect  
involves the challenge of dealing with spectral multicollinearity. The ability to provide robust estimates, retrieval uncertainties and acceptable retrieval processing speed are other important aspects in view of operational processing. Recommendations towards new-generation spectroscopy-based processing chains for operational production of biophysical variables are given.  
\end{abstract}



\section{Introduction} \label{sec:Into}

Quantitative vegetation variable extraction is fundamental to assess the dynamic response of vegetation to changing environmental conditions. Earth observation {sensors} in the optical domain enable the spatiotemporally-explicit retrieval of plant biophysical variables. This data stream has never been so rich as is foreseen with the new generation imaging spectrometer missions. The forthcoming EnMAP \citep{Guanter2015}, HyspIRI \citep{LEE2015}, PRISMA \citep{Labate2009a} and FLEX \citep{Drusch2016a} satellite missions will produce large spectroscopic data streams for land monitoring, which will soon become available to a diverse user community. This upcoming vast data stream  will not only be standardized (e.g. atmospherically-corrected), but will also require reliable and efficient retrieval processing techniques that are accurate, robust and fast. 

Since the advent of optical remote sensing science, a variety of retrieval methods for vegetation attribute extraction emerged.  
Most of these methods have been applied to the data of traditional multispectral sensors \citep{verrelst2015b}, but increasingly they are also applied within imaging spectroscopy studies. This review provides a summary of recently developed methodologies to infer per-pixel biophysical variables from imaging spectroscopy data, covering the visible, near-infrared (NIR) and shortwave infrared spectral regions.  
Essentially, quantification of surface biophysical variables from spectral data always relies on a model, enabling the interpretation of spectral observations and their translation into a surface biophysical variable. Biophysical variable retrievals, as traditionally described in terrestrial remote sensing literature, are grouped into two categories: (1) the statistical (or variable-driven) category; and (2) the physical (or radiometric data-driven) category \citep{Baret2008}. Over the last decade, however, both methodological categories expanded into subcategories and combinations thereof. Exemplary is the increasing number of elements of both categories which have been integrated into \emph{hybrid} approaches. This methodological expansion, therefore, demands for a more systematic categorization. From an optical remote sensing point of view, and in line with an earlier, more general review paper \citep{verrelst2015b}, retrieval methods can be classified in the following four methodological categories:

\begin{enumerate}
\item  \emph{Parametric regression methods}: Parametric methods assume an explicit relationship between spectral observations and a specific biophysical variable. Thus, explicit parameterized expressions are built usually based on some physical knowledge of absorption and scattering properties and statistical relationship between the variable and the spectral response.  
Typically a band arithmetic formulation is defined (e.g., a spectral index) and then linked to the variable of interest based on a fitting function.
\item  \emph{Non-parametric regression methods}: Non-parametric methods directly define regression functions according to information from the given spectral data and associated variable, i.e., they are data-driven methods. Hence, in contrast to parametric regression methods, a non-explicit choice is to be made on spectral band relationships, transformation(s) or fitting function. Non-parametric methods can further be split into linear or nonlinear regression methods.
\item  \emph{Physically-based model inversion methods}: Physically-based algorithms are applications of physical laws establishing photon interaction cause-effect relationships. Model variables are inferred based on specific knowledge, typically obtained with radiative transfer functions.
\item  \emph{Hybrid regression methods}: A hybrid-method combines elements of non-parametric statistics and physically-based methods. Hybrid models rely on the generic properties of physically-based methods combined with the flexibility and computational efficiency of non-parametric nonlinear regression methods.
\end{enumerate}

These categories provide a theoretical framework to organize the myriad of retrieval methods, as well to overview the diversity of published imaging spectroscopy applications based on these methods. However, a few remarks must be considered. One should be aware that the boundaries of these categories are not always clearly defined; for instance, spectral indices are also often used as input into non-parametric methods. 
Another important aspect is that the majority of the here reviewed methods is not exclusively designed for retrieval of biophysical variables. This especially holds for the statistical methods, whereby a regression model is used to link spectral data with a biophysical variable. In optical remote sensing science these methods are commonly applied to map any feasible continuous variable, as well in the domains of snow, water or soil properties (see \citet{Matthews2011}, \citet{Mulder20111} and \citet{Dietz2012} for reviews). Nevertheless, to keep this review comprehensive, it is limited to retrieval methods with applications in the domain of vegetation properties mapping.  
On the other hand, even within these boundaries each of the above methodological categories continue to be expanded  with all kinds of spectroscopic data processing applications  \citep[e.g.][]{Gewali2018}.  
The drivers behind this methodological expansion can be found in the: (1) interminable increase of computational power, (2) the increasing availability and democratizing of spectroscopic data, and (3) 
the steady progress in imaging spectroscopy sensor technology, which produces each time more sensitive sensors. This progress in imaging spectroscopy technology enables to infer each time more subtle and highly dynamic vegetation properties from spectral data. For instance, the forthcoming FLEX mission aims to deliver a portfolio of dynamic plant stress and productivity variables based on, among others, the exploitation of sun-induced chlorophyll fluorescence emitted by terrestrial vegetation \citep{Drusch2016a}. Hence, this underlines the fact that the list of biophysical variables that can be extracted from imaging spectroscopy is not closed, but instead  
continues to grow with ongoing progress in spectrometer technology.  
Consequently, biophysical variables are in this review paper defined as any vegetation property that can be quantified, i.e. any pigments, chemical constituents, structural variables, but also variables related to plant photosynthesis, productivity or diseases.  
Altogether, the drivers behind methodological expansion are not mutually exclusive, but they strengthen each other, which leads to a rapid progress in the development of advanced retrieval methods that goes hand in hand with improved capabilities to quantify a broad diversity of biophysical variables. As will be demonstrated throughout this review, these trends are resulting in an unprecedented richness of imaging spectroscopy mapping applications. 
 
Regardless of the used methodology or the targeted application, the principal characteristic of spectroscopic data lies in their dense information content embedded in a few hundred spectrally narrow bands. Although such spectrally dense data source proved to be beneficial for the majority of targeted mapping applications, a key challenge for many retrieval methods is how to deal with spectral multicollinearity, i.e. band redundancy. Special attention, therefore, will be devoted to address common spectroscopic data processing challenges, and solutions will be given how to overcome them.  
Finally, while imaging spectrometers are so far mostly applied in an experimental context, the developments towards operational systems have manifestly taken off -- and undoubtedly will lead to new directions and possibilities of Earth observation. In view of {getting prepared for these upcoming} 
 global spectroscopic data streams, we will close this review with recommendations about the possibilities of integrating promising retrieval approaches into operational schemes.

\section{Parametric regression methods}  \label{sec:param}

Parametric regression methods have long been the most popular method to quantify biophysical variables in optical remote sensing; and the field of imaging spectroscopy is no exception to that. This simplest way of developing a regression model explicitly determines parameterized expressions relating a limited number of spectral bands with a biophysical variable of interest. The empirical models rely on a selection of bands with high sensitivity towards the variable of interest, typically in combination with subtle spectral features to reduce undesired effects; related to variations of, for instance, other leaf or canopy properties, background soil reflectance, solar illumination and sensor viewing geometry and atmospheric composition \citep[e.g.][]{Verrelst08,Verrelst10}. In the following overview we present common parametric regression methods, which are based on (1) vegetation indices, (2) shape indices, and (3) spectral transformations.  

\vspace{-1.0em}
\begin{figure*}[!ht]
 \centering
	\footnotesize
	\begin{tabular}{c}
 \IG[width=12 cm, trim={0.0cm 1.0cm 0.0cm 0.0cm}, clip]{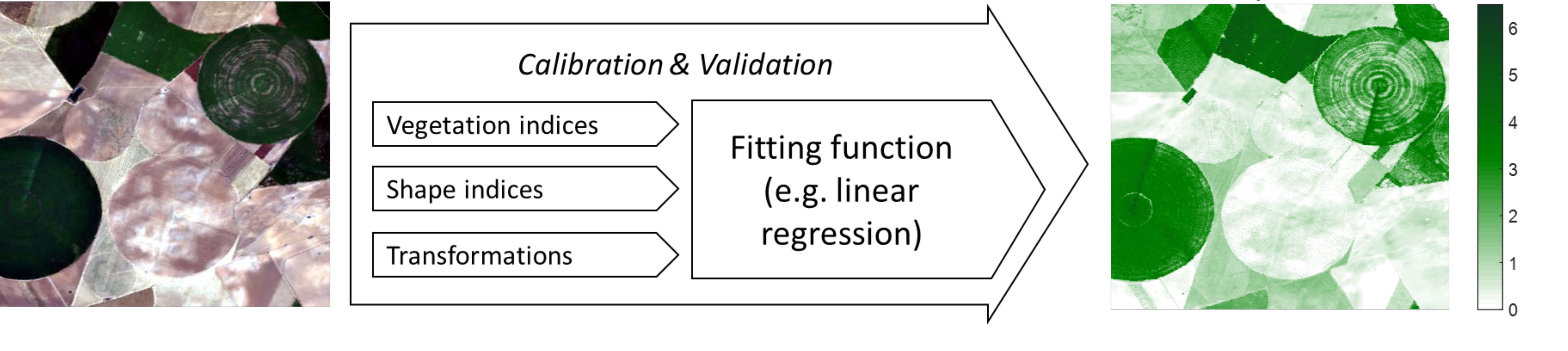} \\

    \end{tabular}\vspace{6pt}
     \caption{Principles of parametric regression. Left: RGB subset of a hyperspectral HyMap image (125 bands) over Barrax agricultural site (Spain). Right: illustrative map of a vegetation property (LAI, m$^2$/m$^2$) as obtained by a 2-band normalized difference index and linear regression. The model was validated with a R$^{2}$ of 0.89 (RMSE: 0.63; NRMSE 10.1\%). It took 0.2 seconds to produce the map using ARTMO's SI toolbox \citep{Rivera2014}. No uncertainty estimates are provided. }
 \label{schematic_overview}
\vspace{-1.5em}
\end{figure*} 

\subsection{Discrete spectral band approaches: vegetation indices}
 
Parametric regression models based on vegetation indices (VIs) are by far the oldest and largest group of variable estimation approaches.  
VIs are defined to enhance spectral features sensitive to a vegetation property, while reducing disturbances by combining some spectral bands into a VI \citep{Glenn2008,Clevers2014}. The main advantage of VIs is their intrinsic simplicity. VI-based methods found their origin in the first applications of broadband sensor satellites. During the pioneering years of optical remote sensing only a small set of spectral bands were available and computational power was limited. It led to a long tradition of the development of simple two-bands, or at most three to four band indices that continues until today \citep[e.g.][]{Kira2016}. 
New possibilities have opened with the advent of imaging spectrometers. Optimized narrowband information extraction algorithms were developed based on adaptations of established index formulations, such as simple ratio, normalized difference (see reviews \citet{Glenn2008,Clevers2014,Xue2017}).  
On the other hand, the possibilities to develop spectral indices based on a few band combinations grew exponentially, and it demanded for more systematic band evaluation methods. 

A popular solution involves correlating all possible band combinations according to established index formulations. For two-band index formulations, such as simple ratio or normalized difference, this approach leads to 2D correlation matrices, which enables to visually identify optimal band combinations \citep[e.g.][]{Thenkabail00,LeMaire04,leMaire08,Atzberger10,Mariotto13,Rivera2014}.  
Subsequently, given all possible combinations permits to select a 'best performing index'. Nevertheless,  while being mathematically simple, this method is not only tedious -- especially when evaluating all possible combinations of more than two bands  -- but also keeps on being restricted to formulations that make use of a few bands only, with at most using three or four bands. Thus, although the approach is systematic, it continues to underexploit the comprehensive information content hidden in the contiguous spectral data. Moreover, when applying this technique in mapping applications making use of imaging spectroscopy, identical best performing spectral band combinations for the same biophysical variable have rarely been reported. This suggests that optimized narrowband VIs are strongly case specific and seem to lack generic capacity \citep{Gonsamo11,Heiskanen13,Mariotto13}.  

More fundamentally, it remains dubious whether relying on transformed data originating from a few discrete bands fully captures the complexity of real world observation conditions as has been observed by a spectroradiometer. Reducing full-spectrum datasets into simple indices formulations intrinsically leads to remaining spectral information left unexploited. 
Accordingly, the following two aspects should be considered to ensure optimized use of VIs in a spectroscopic context: (1) \emph{Band selection}. Spectral indices are mathematical functions based on discrete bands, or at best a subset of full spectral information. Thus, the question arises: how do we assess with high enough accuracy whether the most sensitive spectral bands -- with respect to biophysical variable retrieval -- have been selected? (2) \emph{Formulation}. Enhancing spectral information according to a mathematical transformation should lead to an optimal sensitivity of the spectral signal with respect to the variable of interest. While established formulations such as the simple ratio or normalized difference are commonly used, here the question arises again: how can we be sure whether these linear formulations are the most powerful ones with respect to biophysical variable retrieval? These two questions are almost impossible to resolve considering the unlimited possibilities of  band selections together with designing index formulations.
 Consequently, given their inherent constraints, it can be be concluded that VI-based regression models exploit spectroscopic data suboptimally.

\subsection{Parametric approaches based on spectral shapes and spectral transformations} \label{sec:Param}

Because none of the above few-band indices methods take full advantage of spectroscopic datasets, alternative methods were pursued with the advent of hyperspectral spectroradiometers that allow to exploit specific absorption regions of the reflectance spectrum. It led to the development of so-called \emph{shape} indices and spectral transformation methods. Shape indices, listed below, extract shape-related information from contiguous spectral signatures for a specific spectral region that is then correlated with a biophysical variable.  
These types of parametric methods are therefore exclusively applicable to spectroscopic data. 
The following categories can be identified: 
\begin{itemize}
  \item \emph{Red-edge position (REP) calculations}. Mathematically, the REP inflection point is the position of a wavelength defined as the maximum of the first derivative reflectance between the red and NIR regions, i.e., between 670 and 780 nm \citep{Kanke2016}. The red-edge position is known to be sensitive to multiple biophysical variable variations, both chlorophyll pigments \citep{Delegido11} as well as structural variables, for instance the leaf area index (LAI) \citep{Delegido2013}. Therefore, REP-related methods are typically used to derive canopy chlorophyll content, being the product of LAI and leaf chlorophyll content \citep{Clevers2012,Li2017}. Many mathematical approaches have been proposed to exploit this region as a sensitive indicator, including: (1) high-order curve fitting \citep{Broge01,Clevers2004}; (2) inverted Gaussian models \citep{miller1990,Cho2006,Cho2008}; (3) linear interpolation and extrapolation methods \citep{Cho2008,Tian2011}; (4) Lagrangian interpolation \citep{dawson1998,Pu2003}; (5) rational function application \citep{baranoski2005}; and more recently, (6) a wavelet-based technique \citep{Li2017}. 
  
  \item \emph{Derivative-based indices}. Although several of the above-described methods make use of derivatives, e.g. linear extrapolation \citep{Cho2006} and Lagrangian technique \citep{dawson1998}, the calculation of a derivative does not have to be restricted to the red edge. The derivative of any spectral region can be calculated and transformed into an index \citep{Penuelas94,Elvidge1995,Sims02,zarco2002}. A systematic comparison of first derivative-based indices and conventional indices was performed by \citet{LeMaire04} using the leaf optical model PROSPECT. Interestingly, the authors concluded that derivative-based indices are not necessarily better than conventional and properly elaborated indices.

  \item \emph{Integration-based indices}. Alternatively, some authors proposed to calculate finite integrals of specific spectral regions, typically covering a part of the visible and the red-edge region for LAI or chlorophyll content estimations, into a (normalized) index \citep{Broge01,oppelt2004,Mutanga2005,Malenovsky2006,malenovsky2015,Delegido10}. Likewise, in a recent study of \citet{Pasqualotto2018} this method exploited the water absorption spectral regions to quantify canopy water content.  In these studies, integration-based indices were demonstrated to perform superior to classical vegetation indices, as they exploit more optimally absorption regions embedded in spectroscopic data than indices relying on a reflectance intensity of few individual bands \citep{kovavc2013}. It can be expected that with the upcoming free availability of imaging spectroscopy data more of {this} kind of methods that explicitly exploit absorption features related to foliar constituents and pigments will emerge.  
  \item \emph{Continuum removal}. Whereas the above techniques focus on one or more specific spectral regions, continuum removal is a spectral transformation that can be applied over the full spectrum. This technique normalizes reflectance spectra, allowing comparison of individual absorption features with a common baseline \citep{clark1984}. The continuum removal transformation enhances and standardizes the specific absorption features related to vegetation properties. Continuum removal can be considered as a standard spectroscopic data processing technique and has found its way in various image processing software packages. Spectroscopic examples of  applications include mapping of chlorophyll \citep{Broge01,Malenovsky2013,Malenovsky2017}, numerous studies on mapping nitrogen content \citep{huang2004,mutanga2004,Mutanga2007,schlerf2010,mitchell2012,Yao2015}, foliar water condition \citep{stimson2005}, plant stress \citep{Sanches2014} and grassland biomass \citep{Cho2007,Buchorn2013}. 
  
  \item \emph{Wavelet transform}. Wavelet analysis has been increasingly used to extract information from spectral data, e.g. related to vegetation properties \citep{Rivard2008}. Processing of reflectance spectra with wavelets can be performed as discrete or continuous (CWT) transforms. CWT outputs are directly comparable to the original spectrum and are simple to interpret. In this case, the original spectrum is represented by a set of spectra from small (narrow bandwidth absorption feature and noise) to larger scales (broad features, continuum). By selecting small scale spectra (i.e. discarding the smallest scale, which contains white noise and high scales related to the continuum), the absorption features of the components are enhanced, preserving the spectral information of the original data \citep{Scafutto2016}. Based on the type of wavelet transform, specific bands sensitive to the targeted variable are then selected \citep{Bao2017}. CWT is often compared in spectroscopic studies against spectral indices and was found {to be capable of delivering} stronger correlations, e.g. in the detection of wheat aphid pests \citep{Luo2013}, LAI estimation \citep{Huang2014}, nitrogen content and chlorophyll content estimation \citep{Luo2013, Kalacska2015, He2015} and in amplifying spectral separability of alpine wetland grass species. \citep{Bao2017}.

\end{itemize}

Altogether, correlations based on shape indices and spectral transformations are undoubtedly more sophisticated normalization approaches than traditional spectral indices for exploiting the spectral information embedded in spectroscopic data. Moreover, their relatively simple mathematical formulation ensures fast processing. It seems thus logical that these spectral transformation methods became standard spectroscopy image processing techniques. However, these methods alone provide nothing more than spectral transformations and enhancements. 
When aiming to estimate a biophysical variable, a fitting function -- typically a linear least squares fitting, but also exponential, power and polynomial -- is still required. Yet it remains questionable whether the selected fitting function is the most suitable one. 
Moreover, since parametric approaches are based on relatively simple mathematical definitions -- as opposed to more advanced methods --  no associated uncertainty intervals are provided. Although their strengths lie in their straightforward use and fast processing, with the absence of a per-pixel uncertainty estimate, the performance quality of parametric regression methods as a mapping method is hard to judge. 
Given the surface diversity captured in a single airborne or spaceborne image, and despite a standard validation exercise for a number of pixels, it still  remains unknown how the retrieval quality evolves throughout a complete image. The absence of a quality indicator is, therefore, in our view the main reason why parametric regression methods are not recommendable for operational quantification of biophysical variables.

\begin{figure*}[!h]
 \centering
	\footnotesize
	\setlength\tabcolsep{2pt} 
	\begin{tabular}{ccc}
	\textbf{(a) Spectral indices }& \textbf{(b) REP calculation} & \textbf{(c) Derivative-based indices} \\
 \IG[width=3.8 cm, trim={0.5cm 0.0cm 1.5cm 0.0cm}, clip]{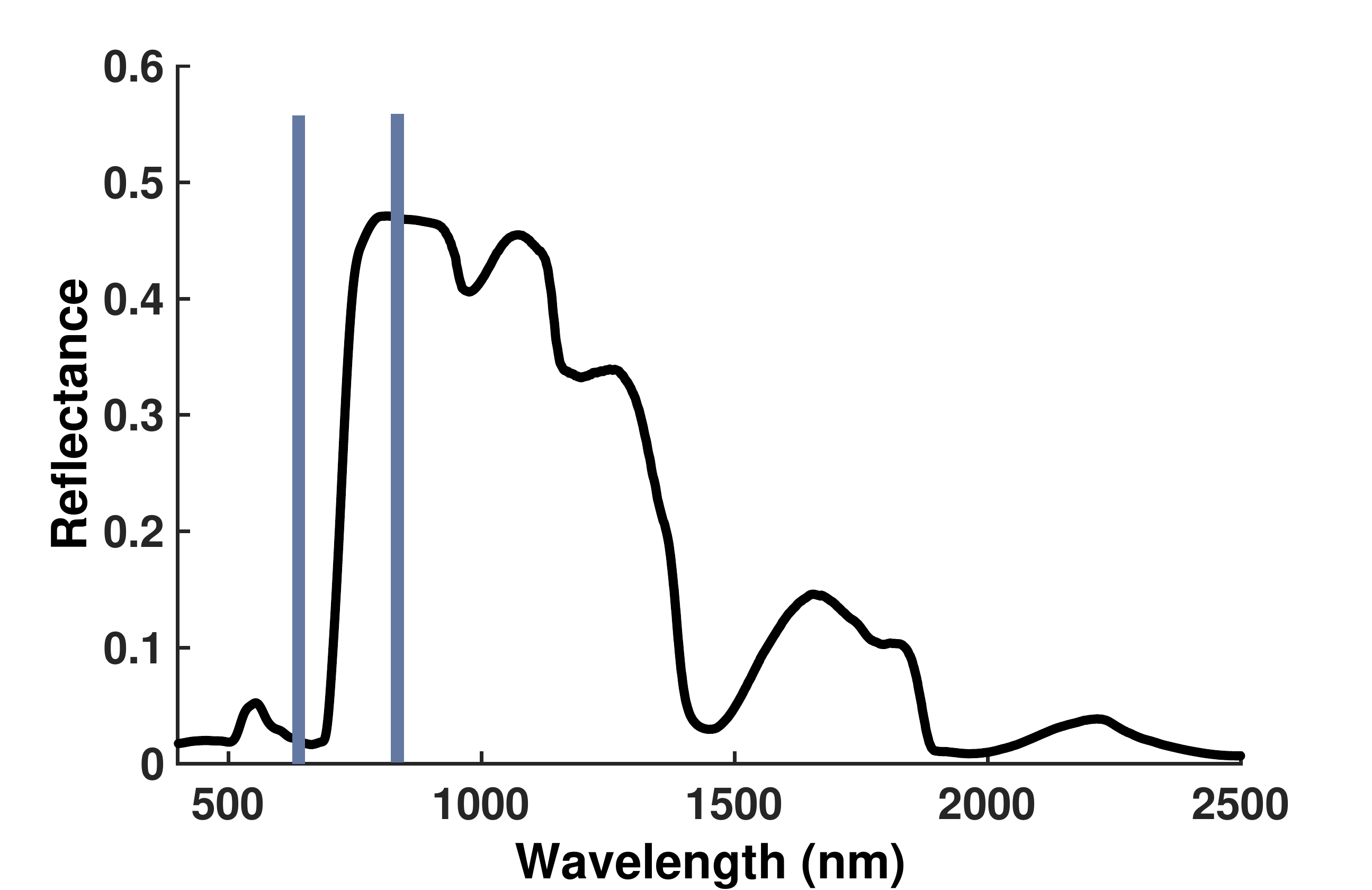} &  
 \IG[width=3.8 cm, trim={0.5cm 0.1cm 1.5cm 0.0cm}, clip]{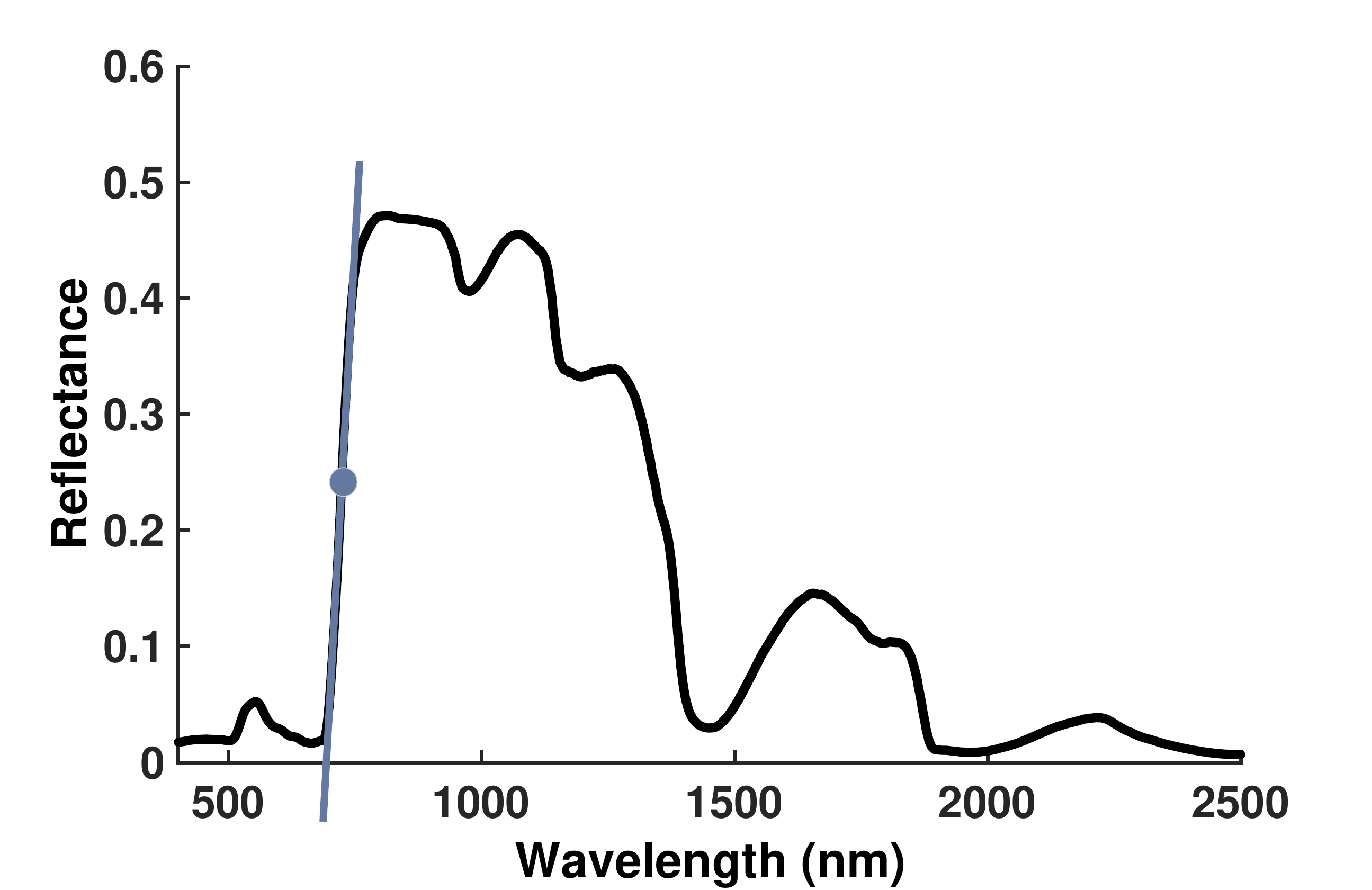} & 
 \IG[width=3.8 cm, trim={0.5cm 0.1cm 0.0cm 0.0cm}, clip]{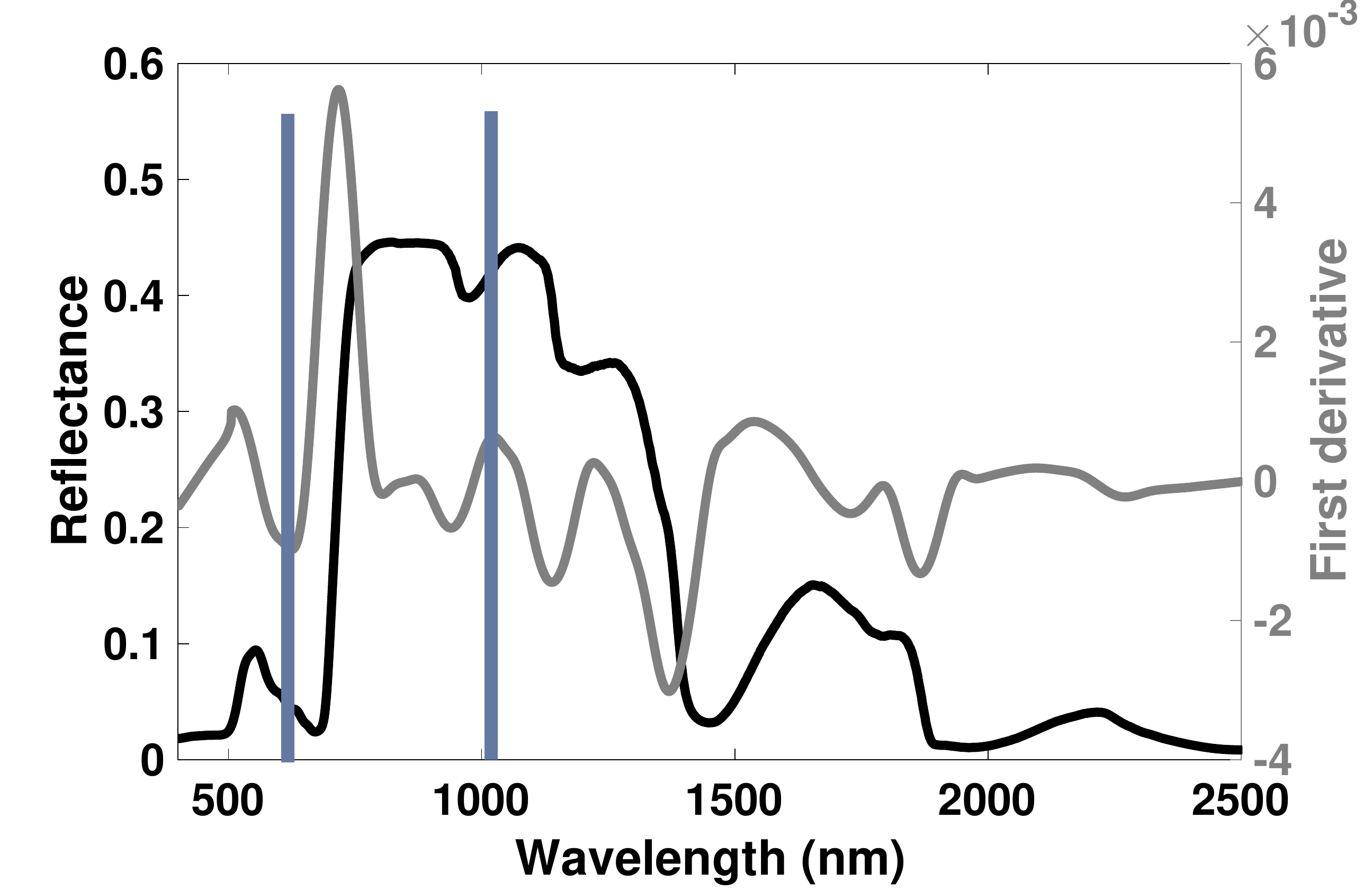} \\ 
\\
\textbf{(d) Integral-based indices} & \textbf{(e) Continuum removal} & \textbf{(f) Wavelet transform} \\
 \IG[width=3.8 cm, trim={0.5cm 0.2cm 1.5cm 0.3cm}, clip]{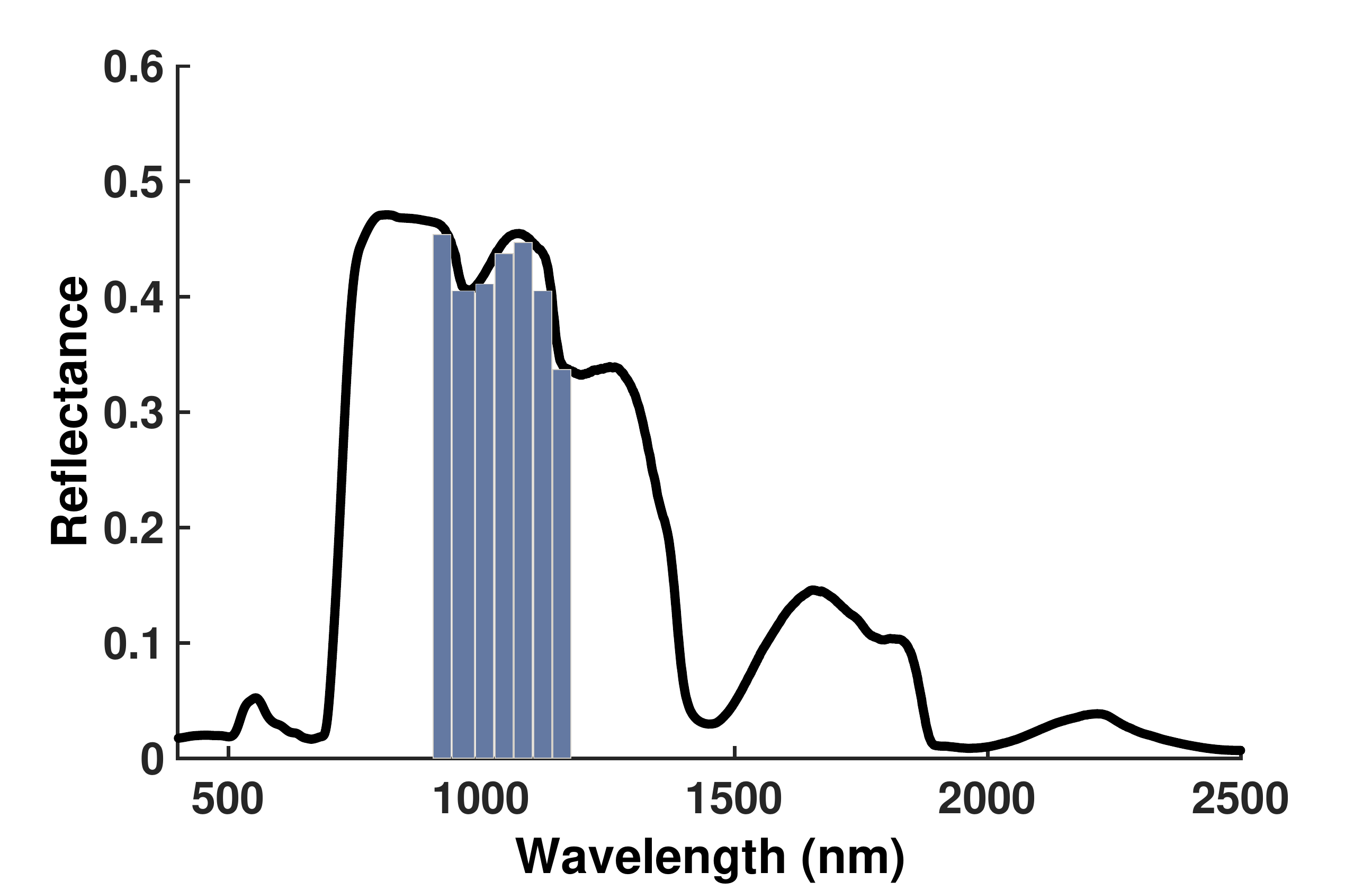} & 
 \IG[width=3.8 cm, trim={0.5cm 0.3cm 1.5cm 0.3cm}, clip]{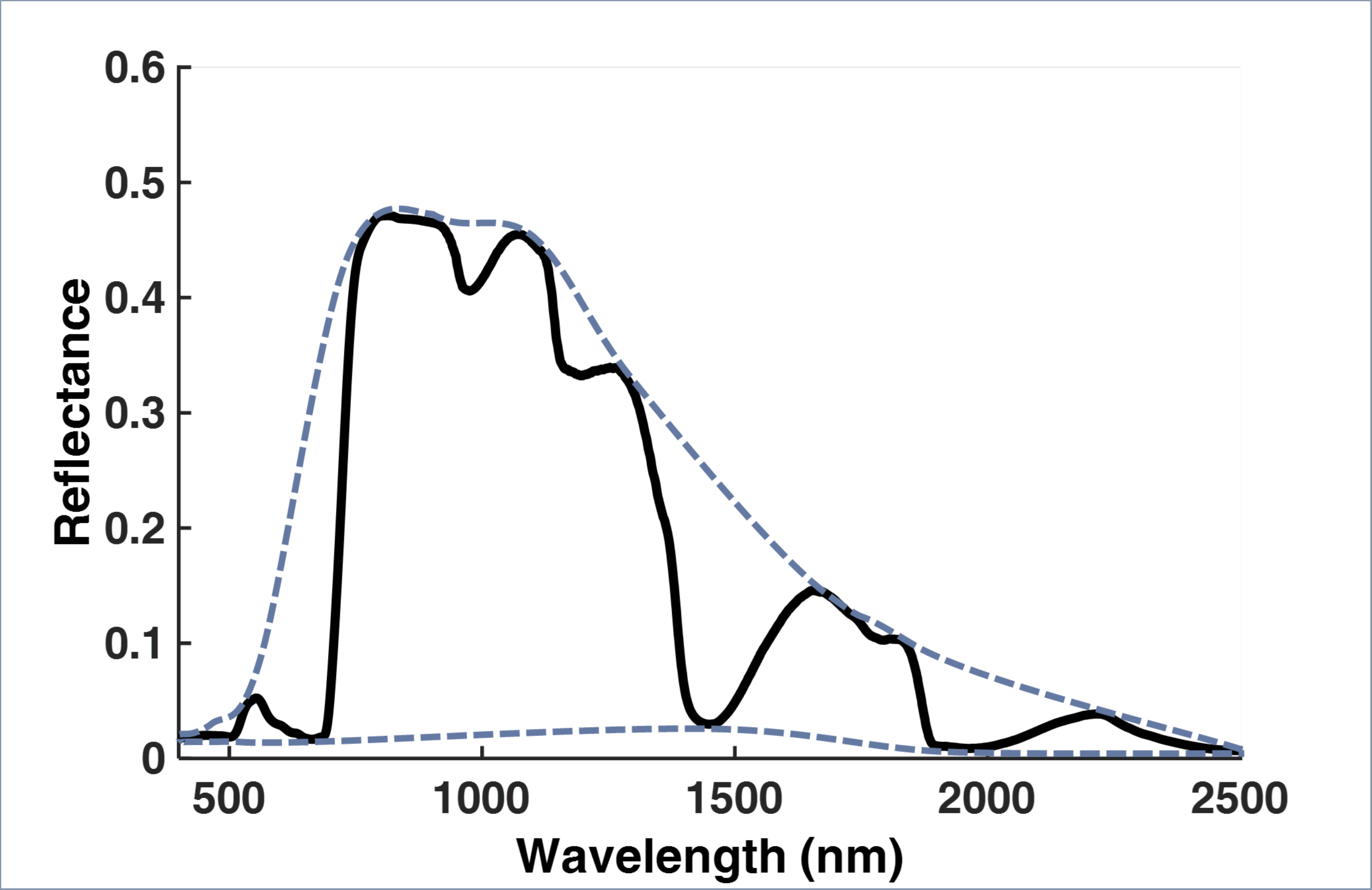} & 
 \IG[width=3.8 cm, trim={0.5cm 0.3cm 1.5cm 0.3cm}, clip]{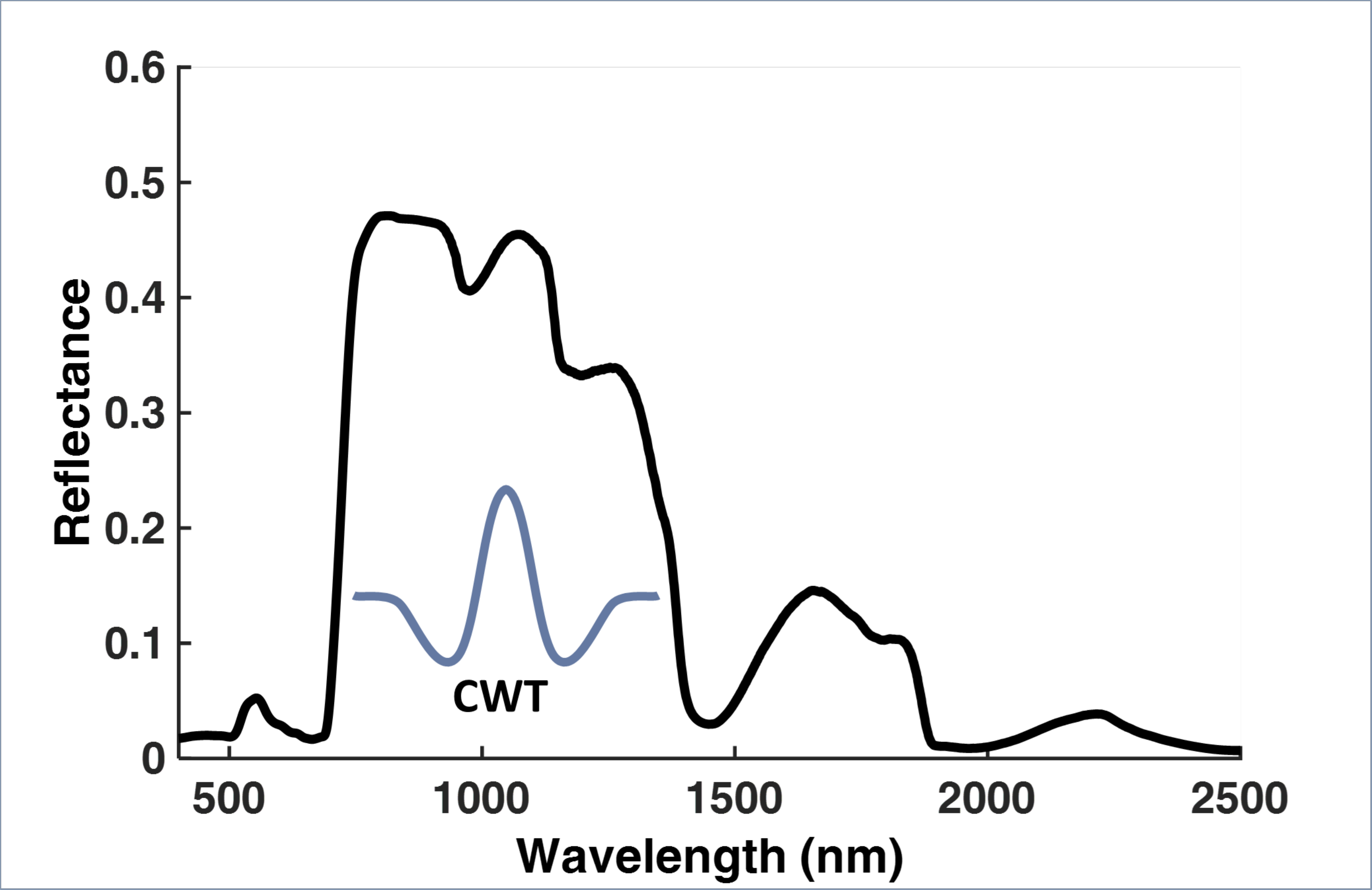} \\ 
    \end{tabular}\vspace{6pt}
     \caption{{Schematic illustrations of parametric regression methods: Spectral indices (a), Red-edge position (REP) calculation (b), derivative-based indices (c), integral-based indices (d) continuum removal (e), and wavelet transform (f). Note that a fitting function is still required to convert transformations towards a biophysical variable.} }
 \label{schematic_overview}
\vspace{-1.5em}
\end{figure*}

\section{Non-parametric regression methods}  \label{sec:Nonparam}

Contrary  to parametric methods, non-parametric methods optimize the regression algorithm by means of an inherent learning phase based on training data. Essentially, the non-parametric model develops weights (coefficients) adjusted to minimize the estimation error of the variables extracted. This means that no explicit parametrization is required, which practically simplifies the model development, but more expert knowledge to understand and execute these models may be required.  
Another important advantage of non-parametric methods is the possibility of training with the full-spectrum information. Hence, an explicit selection of spectral bands or transformations is in principle not required. A flexible model is able to combine different data structure features in a nonlinear manner to conform requirements; however model definition with a too flexible capacity may incur the problem of over-fitting the training dataset. To avoid this pitfall, model weights are defined by jointly minimizing the training set approximation error while limiting the model complexity.  
In view of processing spectroscopic data, a more prevalent problem lies in the so-called \emph{curse of dimensionality} (Hughes phenomenon) \citep{Hughes1968}. Adjacent, contiguous bands carry highly intercorrelated information, which may result in redundant data and possible noise and potentially suboptimal regression performances. As discussed further on, band selection or dimensionality reduction methods that transform the spectral data to lower-dimensional space, while containing the vast majority of the original information can overcome this problem.

\begin{figure*}[!ht]
 \centering
	\footnotesize
	\begin{tabular}{c}
 \IG[width=12 cm, trim={0.0cm 1.0cm 0.0cm 0.0cm}, clip]{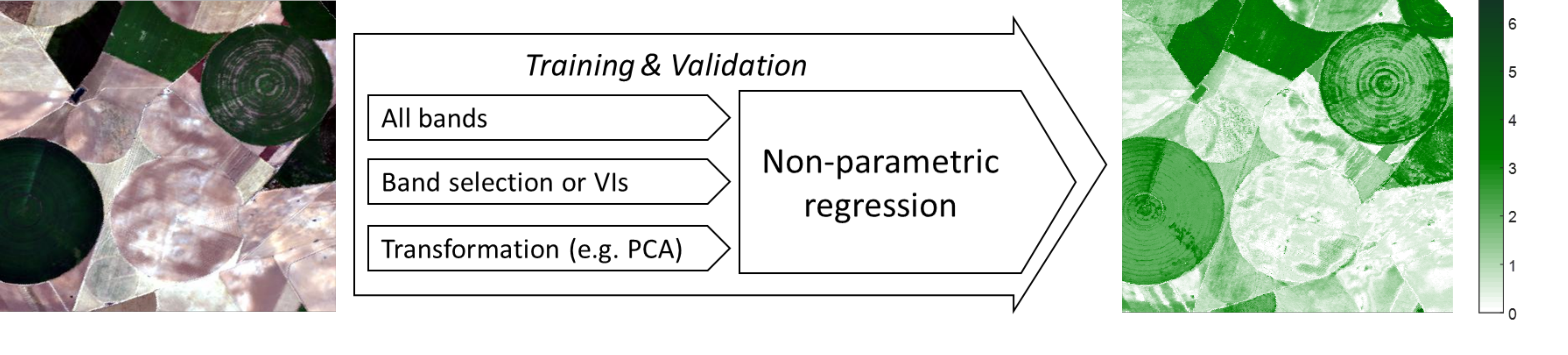} \\

    \end{tabular}\vspace{6pt}
     \caption{Principles of non-parametric regression. Left: RGB subset of a hyperspectral HyMap image (125 bands) over Barrax agricultural site (Spain). Right: illustrative map of a vegetation property (LAI, m$^2$/m$^2$) as obtained by PROSAIL with Gaussian processes regression (GPR). The model was validated with a R$^{2}$ of 0.94 (RMSE: 0.39; NRMSE: 6.3\%). It took 5.7 seconds to produce the map using ARTMO's MLRA toolbox \citep{Caicedo2014}. With GPR also uncertainty estimates are provided (not shown). }
 \label{schematic_overview}
\vspace{-1.5em}
\end{figure*}

\subsection{Linear non-parametric methods}  \label{sec:lin-Nonparam}

Non-parametric regression algorithms that apply linear transformations are attractive because of their fast performance. These methods became standard methods in chemometric and in image processing software packages. 
Multivariable linear regression methods can cope with spectroscopic data and typically rely on the estimation of co-variances. When moving towards spectroscopic data, however, this can become problematic when input data quantity is limited with respect to the dimensionality of the dataset. To alleviate collinearity, often linear non-parametric methods are applied in combination with a dimensionality reduction step. Some methods are even intrinsically based on this principle, i.e. principal component regression (PCR) \citep{Wold1987}, and partial least squares regression (PLSR) \citep{Geladi86}. Common linear non-parametric regression approaches are provided in table \ref{linear_nonpar} and imaging spectroscopy applications are discussed below.

\begin{table}[h] 
\scriptsize
\centering
\caption{Linear non-parametric regression methods applicable to spectroscopic data.}
\label{linear_nonpar}
\begin{tabular}{|>{\raggedright}p{1.5cm}|p{8.6cm}|p{1.1cm}|}
\hline
\textbf{Method} & \textbf{Description}  & \textbf{Ref.}  \\
\hline
Stepwise multiple linear regression (SMLR) & SMLR recursively applies multiple regression a number of times. Each step removes a variable eliciting the weakest correlation. At the end of the recursive process, a variable set is obtained that is optimally explaining the spectral data distribution.   & \citet{draper2014}  \\
\hline
Principal components regression (PCR) & PCR is a regression analysis method based on principal components analysis (PCA) estimating regression coefficients. Solutions from PCR are generated performing linear regression of the most relevant components (called scores) obtained after applying PCA. &  \citet{Wold1987}  \\
\hline
Partial least squares regression (PLSR) & PLSR is similar as PCR but tackles the co-linearity problem differently than PCR. Applying PCR, regression is performed using PCA scores. These projections are obtained using only input patterns, not outputs. In contrast, PLSR builds the regression model on projections obtained using the partial least squares (PLS) approach. It elicits the directions of maximum input-output cross-covariance. Therefore, PLSR takes both input patterns and output variables into account. & \citet{Geladi86} \\
\hline
Ridge (regulated) regression (RR)  &  
RR is the most commonly used method of regularization for ill-posed problems, which are problems that do not have a unique solution. RR deals with co-linearity by allowing a degree of bias in the estimates. Therefore, RR adds a small positive value $\lambda$ to the diagonal elements of the input data covariance matrix. Hence, RR requires finding an optimal value for $\lambda$. Typically, cross-validation is used to reach near optimal values. An important fact about RR is that it enforces the regression coefficients to be lower, but it does not enforce them to be zero. That is, it will not get rid of irrelevant features (bands) but rather minimize their impact on the trained model. & \citet{Geladi86} \\
\hline
Least absolute shrinkage and selection operator (LASSO) & Lasso is an extension built on RR, but with a small twist. It also penalizes the regression coefficients absolute size. By this penalization some of the variable estimates may be exactly zero. The larger the penalty, the more the estimates will tend toward zero. This is a convenient approach to automatically perform feature selection, or to deal with correlated predictors.  & \citet{tibshirani1996} \\ 
\hline
\end{tabular}
\end{table}

On the application side, stepwise multiple linear regression (SMLR) is a classical multivariable regression algorithm commonly applied in chemometrics \citep{Atzberger10}. To evaluate its predictive power, SMLR has been often compared with alternative regression techniques such as PLSR and some studies concluded that PLSR yielded better results when estimating LAI \citep{Darvishzadeh08} and canopy chlorophyll content \citep{Atzberger10}. Also \citet{Ramoelo2011} compared both regression algorithms to estimate foliar nitrogen and phosphorus in combination with continuum removal using field spectrometry. By estimating canopy nitrogen, \citet{miphokasap2012} demonstrated that the model developed by SMLR led to a higher correlation coefficient and lower errors than model applications based on narrowband VIs. This suggests that non-parametric (full-spectrum) models tend to be more powerful than parametric models. Likewise,  \citet{Yi2014} compared SMLR with PLSR and spectral indices for carotenoid estimation in cotton  and concluded that best estimations were obtained with PLSR. Likewise, SMLR was compared with PLSR and (nonlinear) machine learning regression algorithms for estimating leaf nitrogen content \citep{Yao2015}. Because of their enhanced flexibility, it may not be a surprise that the nonlinear methods outperformed SMLR and PLSR. This was also observed by various similar studies, as will be addressed in section \ref{sec:nonlin-Nonparam}.  

PCR seems to be more effective in the conversion of spectroscopic data into the estimation of vegetation properties, because the PCA-based dimensionality reduction method is embedded in the method in combination with a linear regression function. Hence, by converting the spectral data to a lower dimensional space automatically overcomes the band redundancy problem. This method has been improved with PLSR, where the projections are optimized in view of the regression. It is, therefore, not a surprise that only few spectroscopic studies examined the predictive power of PCR. Those studies compared PCR against PLSR or against VIs \citep{Atzberger10,Fu2012,Caicedo2014,Marshall2014,Wang2017a}. Although PCR generally outperformed VIs in explaining variability of a vegetation attribute, in all cases PLSR or any other non-parametric method overran PCR. 

PLSR found its way in a broad diversity of imaging spectroscopy applications, especially in the mapping of biochemicals, pigments and vegetation density properties. For instance, PLSR was used in several spectroscopic studies applied to estimate foliage nitrogen content \citep{Hansen2003,Coops2003,huang2004}. Also \citet{Gianelle2007a} used PLSR to derive grassland phytomass and its total (percentage) nitrogen content from spectroscopic data. Similarly, \citet{Cho2007} and \citet{Im2009} applied PLSR to estimate a diversity of grass and crop biophysical variables (LAI, stem biomass and leaf nutrient concentrations), and \citet{Ye2007} applied PLSR for yield prediction purposes. Beyond individual vegetation attributes, PLSR was recently used to predict landscape-scale fluxes of net ecosystem exchange (NEE) and gross primary productivity (GPP) across multiple timescales \citep{Matthes2015}, and also for the estimation of floristic composition of grassland ecosystems \citep{Harris2015,Roth2015,Neumann2016}.
At the same time, thanks to its PLS-vectors, PLSR is also increasingly applied for band sensitivity analysis of spectroscopic datasets in view of the targeted application \citep[e.g.][]{Li2014,Feilhauer2015,Neumann2016,Kiala2016,Kira2016}. 
Various experimental studies demonstrated the superior predictive power of PLSR as opposed to VIs for the prediction of multiple vegetation properties, including above-ground biomass, LAI, leaf pigments (chlorophyll, carotenoids), GPP and NEE fluxes, leaf rust disease detection and nutrients concentration (nitrogen and phosphorus concentrations)  
\citep{Hansen2003,Dreccer2014,Capolupo2015,Matthes2015,Foster2017,Yue2017,Wang2017}. However, when compared against machine learning methods, then PLSR no longer appeared to be top performing \citep{Yao2015,Wang2015,Kiala2016,Ashourloo2016}. As will be addressed in section \ref{sec:nonlin-Nonparam}, this is due to the nonlinear transformation conducted in machine learning methods.

Other linear non-parametric regression methods, such as ridge regression (RR) and LASSO, hardly made it into applications for vegetation properties mapping. Yet a few spectroscopic examples are worth mentioning. For instance, \citet{Addink2007} used RR to map LAI and biomass, and more recently \citet{Bratsch2017} applied LASSO to estimate above-ground biomass quantities among different plant tissue type categories in Alaska. In another biomass estimation study, both RR and LASSO were compared against PLSR \citep{lazaridis2010} and also random forests \citep{zandler2015}. Interestingly, RR and LASSO appeared to be top performing. One may, therefore, wonder why these techniques have not been applied more often. 
On the other hand, these linear methods are increasingly replaced by their nonlinear counterparts. For instance, RR has been replaced by kernel ridge regression (KRR) \citep{Suykens1999}, and also PLSR has been redesigned into a kernel version, i.e. the KPLSR, which proved to be more powerful than PLSR for chlorophyll concentration estimation \citep{Arenas2008}. The family of kernel methods is addressed in section \ref{sec:nonlin-Nonparam}.  That none of these linear methods deliver uncertainty estimates is another drawback. Similar as in case of parametric regression, without uncertainty estimates it remains questionable whether these methods can deliver consistent mapping quality throughout a complete image, or are applicable to other images in space and time.

\vspace{-1.0em}
\begin{figure*}[!h]
 \centering
	\footnotesize
	\setlength\tabcolsep{1pt} 
	\begin{tabular}{ccc}
	\textbf{(a) PCR} & \textbf{(b) PLSR} & \textbf{(c) RR \& LASSO} \\
 \IG[width=2.9 cm, height=3.3cm, trim={0.0cm 0.0cm 0.0cm 0.0cm}, clip]{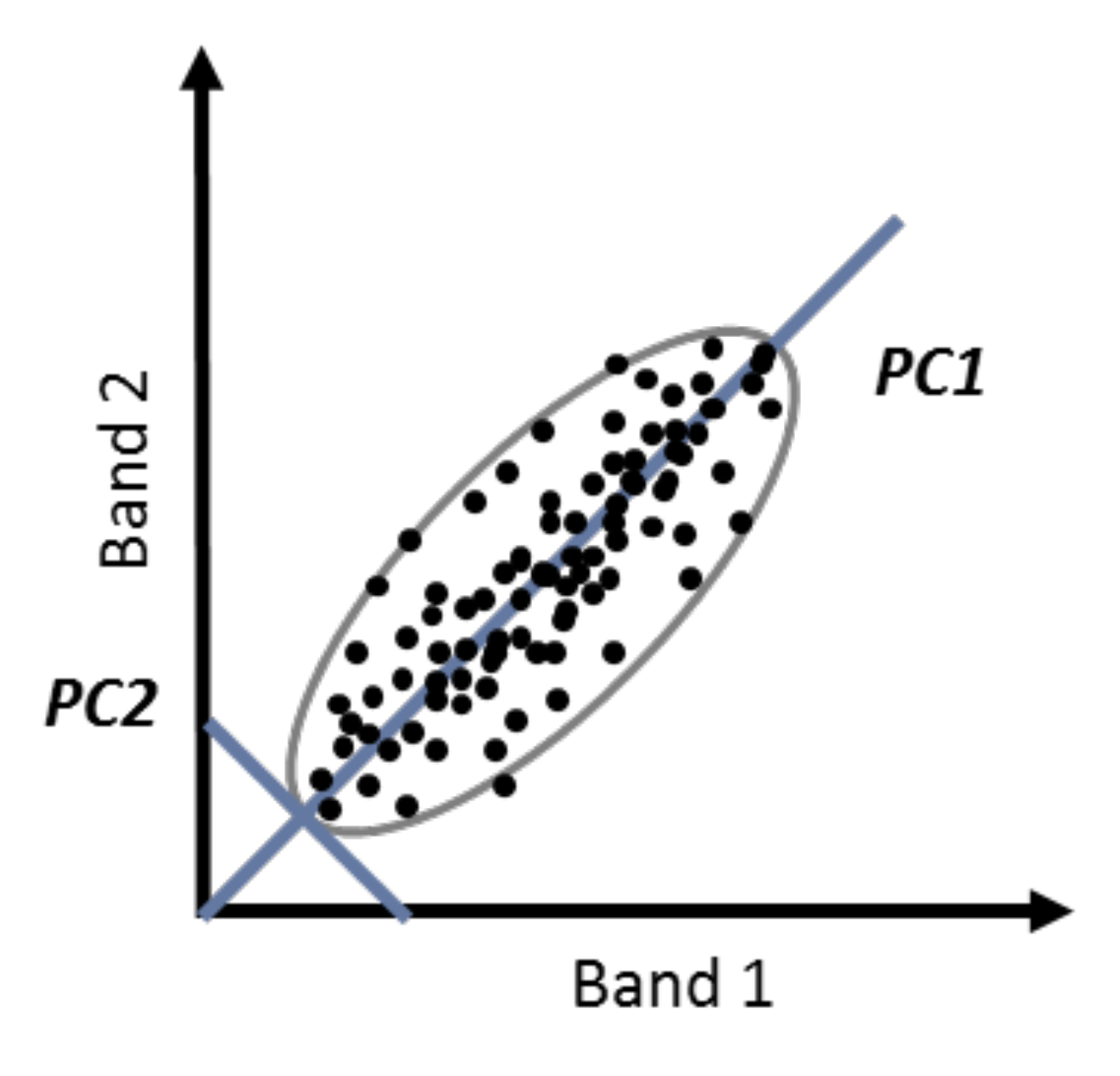} &  
 \IG[width=4.1 cm, height=3.3cm, trim={0.0cm 0.0cm 0.0cm 0.0cm}, clip]{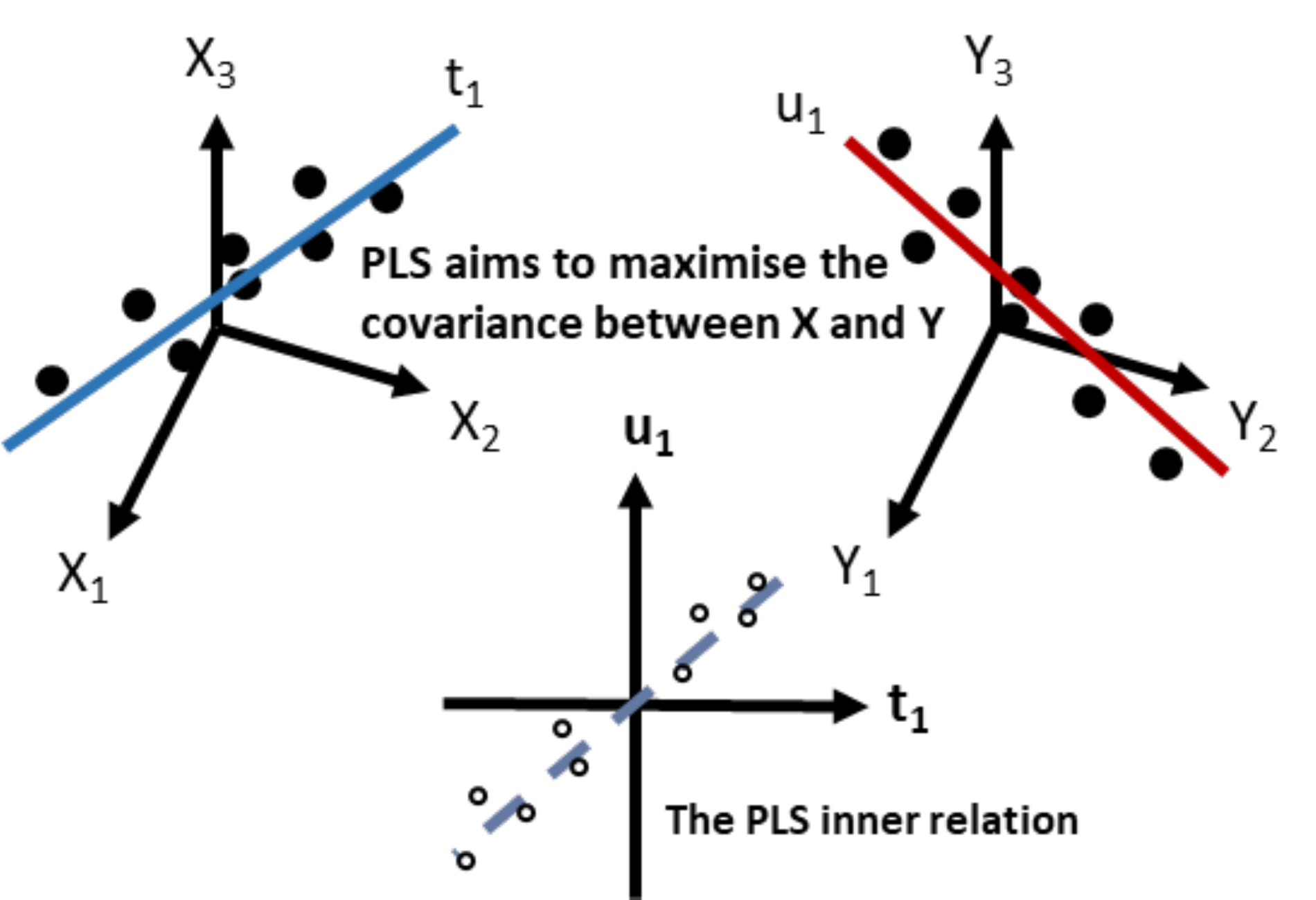} & 
 \IG[width=5.4 cm, height=3.3cm,  trim={0.0cm 0.0cm 0.0cm 0.0cm}, clip]{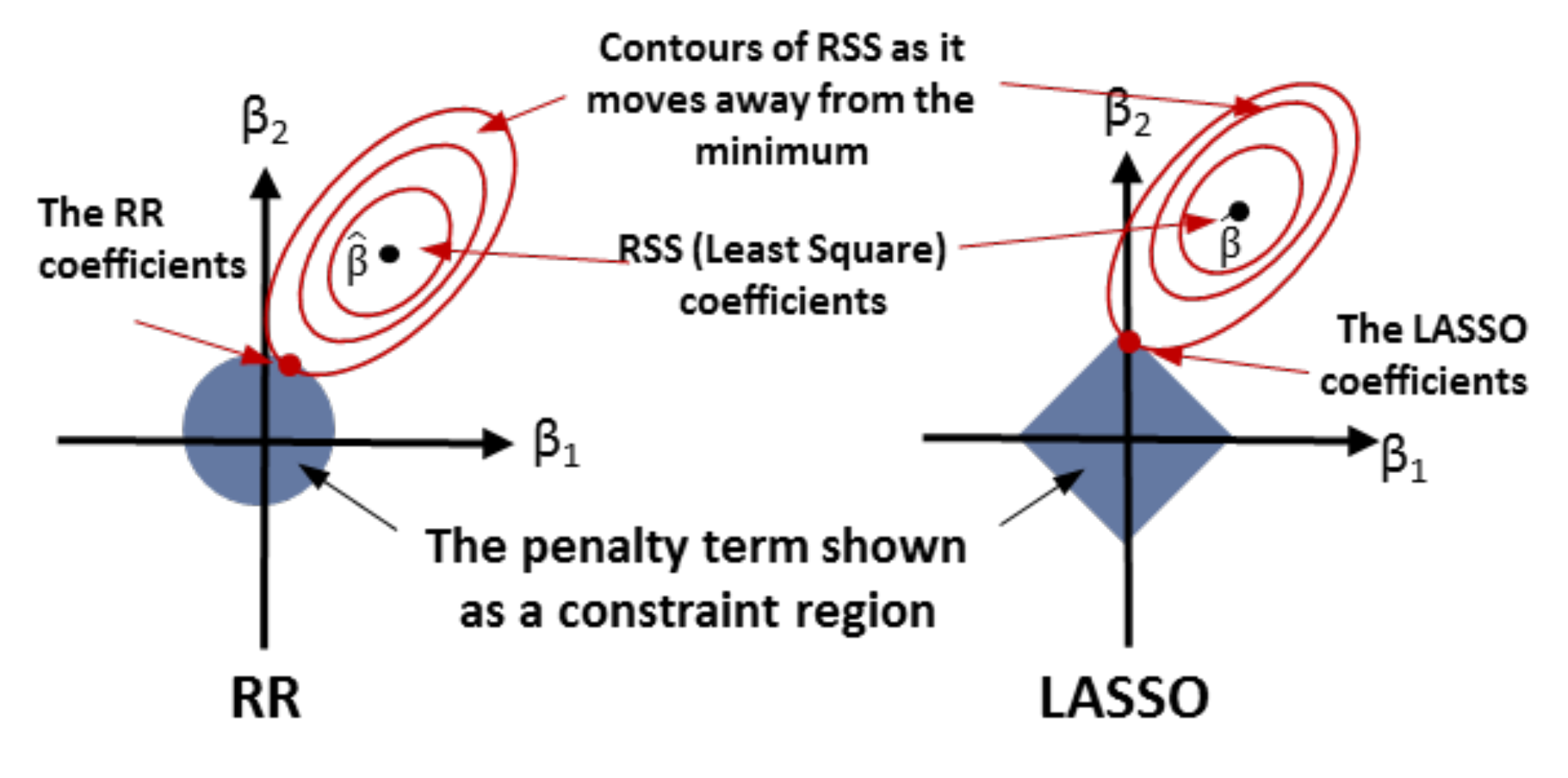} \\ 
    \end{tabular}\vspace{6pt}
     \caption{{Schematic illustrations of principal component (PC) (a), partial least squares (PLS) (B), ridge regression and LASSO (c). PC and PLS is combined with a linear regression model.} }
 \label{schematic_overview}
\vspace{-1.5em}
\end{figure*}

\newpage
\subsection{Nonlinear non-parametric models}  \label{sec:nonlin-Nonparam}

When advancing beyond linear transformation techniques, a diversity of nonlinear non-parametric models has been developed during last few decades. These methods, also referred to as machine learning regression algorithms, apply nonlinear transformations.  
An important methodological advantage is their capability to capture nonlinear relationships of image features without explicitly knowing the underlying data distribution. Hence, they are developed without assuming a particular probability density distribution, which is the reason why they work well with all kinds of data types. 
Machine learning methods also offer the possibility to incorporate a prior knowledge and the flexibility to include different data types into the analysis. In principle they are perfectly suited to process spectroscopic data. In the following sections, examples of the families of (1) decision trees, (2) artificial neural networks and (3) kernel-based regression are explained.

\subsubsection{Decision trees}

\begin{table}[h!]
\scriptsize
\centering
\caption{Decision tree regression methods applicable to spectroscopic data.}
\label{dec_tree}
\begin{tabular}{|>{\raggedright}p{1.5cm}|p{8.6cm}|p{1.1cm}|}
\hline
\textbf{Method} & \textbf{Description}  & \textbf{Ref.}  \\
\hline
Decision trees (DT) & DT learning is based on decision tree predictive modeling. A decision tree is based on a set of hierarchical connected nodes. Each node represents a linear decision based on a specific input feature.     & \citet{Breiman1984} \\
\hline
Boosted trees (BoT) & BoT Incrementally builds an ensemble by training each new instance to emphasize the training instances previously mis-modeled. &  \citet{Friedman2000}  \\
\hline
Bagging trees (BaT) & BaT  an early ensemble method based on building multiple decision trees by iteratively replacing resampled training data and voting for the decision trees leading to a consensus prediction. & \citet{Breiman1996} \\
\hline
Random Forests (RF) & RF is a specific type of BaT that in constructs a collection of decision trees with controlled variance.   & \citet{Breiman2001d} \\
\hline
\end{tabular}
\end{table}

Decision tree algorithms use a branching method to illustrate every possible outcome of a decision (for examples see Table \ref{dec_tree}). They are more frequently applied in classification than in regression. Only a few decision tree feasibility studies dealing with imaging spectroscopy data are presented in the scientific literature \citep[e.g.][]{Im2009} most likely because boosted and bagging trees hardly found their way to regression applications. They might be considered as obsolete with the improvements introduced into random forests (RF), which is essentially a specific type of bagging trees. {RF builds an ensemble of individual decision trees working with different subsets of features (bands) and eventually different training data points both selected randomly, from which a final prediction is made using particular combination schemes. RF can handle a large number of training samples, does not suffer from overfitting and is robust to outliers and noise \citep{Belgiu2016}, which makes it an attractive method for spectroscopic mapping applications.} 
RF has recently been made available in various software packages and proved to be a competent regression algorithm. It therefore comes as no surprise that RF gained rapid popularity in imaging spectroscopy mapping of a diverse range of vegetation attributes, including biomass \citep{Adam2014,Vaglio2014}, canopy nitrogen \citep{Li2014a} and as indicator of plant species composition \citep{Feilhauer2017}. 
Some of these studies have compared RF with support vector regression (SVR) or neural networks, but no strong preference towards one or the other method was found, which suggests that all three methods are competitive \citep{Han2016,Pullanagari2016}.
However, just like other machine learning regression methods, RF can face difficulties coping with the collinearity of the spectroscopic data \citep{RiveraCaicedo2017}. 
To overcome this problem, RF is often used in combination with sensitive bands or simple transformations in the form of VIs that are known to be sensitive to the targeted vegetation property \citep{Adam2014,Liang2016,Han2016}. Alternatively, RF is inherently able to identify sensitive spectral bands, and selection of only those sensitive bands can subsequently improve the regression model \citep{Feilhauer2015,Balzarolo2015}.  
Whether applying a band selection method is the most successful strategy, however, remains an open question. Rather than seeking for optimized individual bands, a more elegant solution may lie in firstly applying dimensionality reduction method, and then inputting the features of the lower-dimensional space (i.e., components) into the decision tree \citep{RiveraCaicedo2017}.

\subsubsection{Artificial neural networks }

\begin{table}[h!]
\scriptsize
\centering
\caption{Artificial neural network regression methods applicable to spectroscopic data.}
\label{ANN_methods}
\begin{tabular}{|>{\raggedright}p{1.5cm}|p{8.6cm}|p{1.1cm}|}
\hline
\textbf{Method} & \textbf{Description}  & \textbf{Ref. } \\
\hline
Artificial neural networks (ANN) & ANNs in their basic form are essentially fully connected layered structures of artificial neurons (AN). An AN is basically a pointwise nonlinear function (e.g., a sigmoid or Gaussian function) applied to the output of a linear regression. ANs with different neural layers are interconnected with weighted links. The most common ANN structure is a feed-forward ANN, where information flows in a unidirectional forward mode. From the input nodes, data pass hidden nodes (if any) toward the output nodes.  & \citet{Haykin1999} \\
\hline
Back-propagation ANN (BPANN) & The he basic type of neural network is multi-layer perceptron, which is feed-forward back-propagation ANN.  BPANN consists of 2 steps: 1) feed forward the values, and 2) calculate the error and propagate it back to the earlier layers. So to be precise, forward-propagation is part of the backpropagation algorithm but comes before back-propagating. This is the most common used algorithm when referring to ANN. In many papers using ANN these standard designs are not explicitly mentioned. &  \citet{Haykin1999}  \\
\hline
Radial basis function ANN (RBFANN) & RBFANN is a type of ANN that uses non-linear radial basis functions (RBF) as activation functions in the hidden layer. The output of the network is a linear combination of RBFs of the inputs and neuron parameters.  & \citet{broomhead1988} \\
\hline
Recurrent ANN (RANN) & A RANN is a type of ANN that make use of sequential information by introducing loops in the network.  & \citet{Hochreiter1997}  \\
\hline
Bayesian regularized ANN (BRANN) & BRANNs are more robust than standard BPANNs and can reduce or eliminate the need for lengthy cross-validation. Bayesian regularization is a mathematical process that converts a nonlinear regression into a "well-posed" statistical problem in the manner of a ridge regression.  & \citet{Burden1999} \\

\hline
\end{tabular}
\end{table}

Artificial neural networks (ANNs) methods are listed in table \ref{ANN_methods}. Since the early 90s, feed-forward and back-propagation ANNs thrived in all kinds of mapping applications, including vegetation properties mapping \citep{Paruelo1997,Francl1997,Kimes1998}. Their strengths lie in their adaptability that can lead to excellent performances. The superiority of ANNs in vegetation properties mapping compared to parametric models (e.g. those based on VIs) has been demonstrated repeatedly in experimental studies \citep{Uno2005,Wang2013,Malenovsky2013,Kalacska2015}. 
Examples of successful spectroscopic applications include the estimation of foliage nitrogen concentrations  \citep{huang2004} and LAI \citep{Jensen2012,Neinavaz2016}. In both cited studies,  ANN outperformed other linear non-parametric models (e.g. PLSR).
Alternative powerful structures involve RBFANNs, BRANNs and RANNs (for explanation see  table \ref{ANN_methods}). Although these advanced ANNs have been primarily used for classification applications, only recently they were explored to map vegetation properties from spectroscopic data \citep{Wang2013,Chen2015,Feng2016,Pocas2017}. Some of these studies mention the superiority of these advanced ANN designs as compared to standard ANN designs or other machine learning approaches in estimating vegetation properties \citep{Du2016,Pham2017,Li2017}.  

Applying ANNs to spectroscopic data, nonetheless, can be quite challenging due to the multicollinearity. Feeding many bands into an ANN requires a complex design and consequently a long training time. 
Just as with decision trees, a popular approach is applying a band selection or the calculation of several sensitive VIs or shape indices such as red edge position that are then entered either individually or as a combination into the ANN. Various of these band selection studies investigated combinations of VIs that led to best prediction models \citep{Schlerf2006,Mutanga2007,Jia2013,Chen2015,Liang2015b,Feng2016,Pocas2017}.  
As discussed before, it remains questionable whether the selected indices preserve a maximum amount of useful information. In contrary, when compressing the spectral data using dimensionality reduction methods into a lower-dimensional space, then it is ensured that a maximum amount of spectral information is preserved. 
This approach was applied e.g. to assess corn yield \citep{Uno2005} and phosphorus and nitrogen concentrations \citep{Knox2011}. It is therefore not surprising that a study comparing PCA vs. indices inputted into ANNs concluded that the PCA-ANN design outperformed VI-ANN designs \citep{Liu2017}. 
Moreover, given that only linear transformations are applied in PCA, it may even be that more adaptive dimensionality reduction methods yield superior accuracies when combined with ANN, e.g. partial least squares (PLS), or in the field of nonlinear kernel-based dimensionality reduction methods, e.g. kernel PCA (KPCA) or kernel PLS (KPLS). To ascertain this hypothesis, PCA was compared against 10 alternative dimensionality reduction methods in combination with ANN to carry out LAI estimation. As expected, various alternative dimensionality reduction methods outperformed PCA in developing accurate models (e.g., PLS, KPLS, KPCA) \citep{RiveraCaicedo2017}.

\subsubsection{Kernel-based machine learning regression methods}

\begin{table}[h]
\scriptsize
\centering
\caption{Kernel-based regression methods applicable to spectroscopic data.}
\label{kernel}
\begin{tabular}{|>{\raggedright}p{1.5cm}|p{8.6cm}|p{1.1cm}|}
\hline
\textbf{Method} & \textbf{Description}  & \textbf{Ref.} \\
\hline
Support vector regression (SVR) & The Support Vector Machine (SVM) is a supervised machine learning technique that was invented in the context of the statistical learning theory. It was not until the mid-90s that an algorithm implementation of the SVM was proposed with the introduction of the kernel trick and the generalization to the non-separable case. SVR is built on the principle of SVM: a non-linear function is learned by linear learning machine mapping into high dimensional kernel induced feature space. The capacity of the system is controlled by parameters that do not depend on the dimensionality of feature (bands) space.  & \citet{Vapnik1997b} \\
\hline
Kernel ridge regression (KRR) & KRR combines RR with the kernel trick. It thus learns a linear function in the space induced by the respective kernel and the data. For non-linear kernels, this corresponds to a non-linear function in the original space. The form of the model learned by KR is identical to SVR. However, different loss functions are used: KRR uses squared error loss while SVR  uses  $\epsilon$-insensitive loss combined with RR regularization.  &  \citet{Suykens1999}  \\
\hline
Gaussian process regression (GPR)  & GPR is based on Gaussian processes (GPs), which generalize Gaussian probability distributions in a function's space. A GP is stochastic since it describes the properties of functions. As in Gaussian distributions, a GP is described by its mean (which for GPs is a function) and covariance (a kernel function). This represents an expected covariance between function values at a given point. Because a GPR model is probabilistic, it is possible to compute the prediction intervals using the trained model. & \citet{Rasmussen2006} \\
\hline
\end{tabular}
\end{table}

Kernel-based regression methods solve nonlinear regression problems by transferring the data to a higher-dimensional space by a kernel function (Table \ref{kernel}). {The flexibility offered by kernel methods allows to transform almost any linear algorithm that can be expressed in terms of dot products, while still using only linear algebra operations. Kernel methods provide a consistent theoretical framework for developing nonlinear techniques and have useful properties when dealing with a low number of (potentially high dimensional) training samples, and outliers and noise in the data \citep{Gomez2011,Tuia2018}. Given these attractive properties, kernel-based regression methods seem perfectly suited to extract nonlinear information related to vegetation properties from imaging spectroscopy data. }  
Developed in the mid-90s, among the most popular kernel-based method for classification purposes involves SVMs. Its regression version (SVR) gained popularity for the retrieval of continuous vegetation attributes from imaging spectroscopy data in the last decade.   
Examples include plant height, leaf nitrogen content, and leaf chlorophyll content \citep{Karimi2008,Yang2011a}. A multi-output version of SVR was presented by \citet{Tuia2011}, with LAI, leaf chlorophyll content and fractional vegetation content being simultaneously estimated.
Recently, SVR was used for processing spectroscopic images of sub-decimeter spatial resolution as acquired by low-altitude unmanned aircraft system to infer Antarctic moss vigour \citep{Malenovsky2017}. Yet just as with the other advanced regression methods,  SVR face the same difficulties of coping with multicollinearity. Therefore, SVR has been commonly applied in combination with specific spectral subsets or VIs \citep{Marabel2013,Lin20139}, or with wavelet transforms \citep{He2015}. To deal with spectroscopic band redundancy, an advantage of SVR is that it allows band selection (analogous as PLSR and RF), which in principle allows the development of more optimized models \citep{Feilhauer2015}.  
On the other hand, it is likely that the combination with dimensionality reduction methods will lead to more powerful models \citep{RiveraCaicedo2017}.
To assess its predictive power, various spectroscopic studies compared SVR against similar methods such as SMLR or PLSR, although some band selection method appeared to be essential \citep{Wang2015,Yao2015,Kiala2016}. Conversely, when comparing SVR against other machine learning methods such as RF or GPR, then SVR no longer excelled \citep{Pullanagari2016}.

Kernel ridge regression (KRR) emerged as one of the promising upcoming kernel-based regression methods, although only a few spectroscopic studies have used it. For instance, \citet{Wang2011} compared KRR with linear non-parametric methods (multiple linear regression and PLSR) for LAI estimation. The authors concluded that KRR yielded the most accurate estimates. Also \citet{Peng2011a} used KRR for the detection of chlorophyll content. Apart from these two studies, the spectroscopy vegetation community may not yet be familiar with this method. Solely \citet{Caicedo2014} had compared KRR against other machine learning algorithms applied to CHRIS (62 bands) and HyMap (125 bands) spectroscopic data for LAI mapping. In that study, KRR not only proved to be a very competitive regression algorithm, but also proved to be extremely fast. This is due to its relatively simple design that requires only one hyperparameter to be tuned. Because of its simplicity, another advantage is that KRR is  capable to deal with collinearity; the method can cope with thousands of contiguous bands. In fact, in the dimensionality reduction comparison study tested with simulated (2100 bands) and HyMap data \citep{RiveraCaicedo2017}, KRR was the only regression method where dimensionality reduction methods did not lead to improvements as compared to using all bands. In conclusion, KRR emerged as an attractive regression method due to its competitive performance, fast processing and easiness to deal with spectroscopic data.

From all machine learning regression algorithms, probably the most exciting one is Gaussian process regression (GPR). Contrary to other methods, the training phase in GPR takes place in a Bayesian framework, leading to probabilistic outputs \citep{Rasmussen2006,CampsValls16grsm}. GPR applied to spectroscopic data started only recently, e.g. for airborne HyMap mapping of leaf chlorophyll content  \citep{Verrelst2013c}, and for spaceborne CHRIS mapping of leaf chlorophyll content, LAI  and fractional vegetation content \citep{Verrelst2012}. Of interest is that along with these maps also maps of associated uncertainties (prediction intervals) were provided. Also with an Airborne Hyperspectral Scanner \citep{roelofsen2014} applied GPR to map salinity, moisture and nutrient concentrations that in turn were used as inputs for plant association mapping. In the \citet{Caicedo2014} comparison paper, GPR  outperformed the majority of other tested machine learning methods for the prediction of leaf chlorophyll content and LAI from various spectroscopic datasets. Similarly, \citet{Ashourloo2016} concluded that GPR yielded most accurate leaf rust disease detection as compared to VIs, PLSR and SVR. However, GPR is no exception in suffering from radiometric collinearity when many bands are included; and related spectroscopic studies demonstrated that results can be further improved by combining GPR with band selection  \citep{Verrelst2016GPRBAT} or with dimensionality reduction methods \citep{RiveraCaicedo2017}. 
At the same time, alternative GPR versions continue to be developed within the machine learning community. For instance, \citet{Lazaro-Gredilla2013} refined the GPR method by proposing a non-standard variable approximation allowing for accurate inferences in signal-dependent noise scenarios. The so-called variational heteroscedastic GPR (VHGPR) appears to be an excellent alternative for standard GPR, which was demonstrated on a CHRIS dataset where VHGPR outperformed GPR in leaf chlorophyll content estimation.

\vspace{-1.0em}
\begin{figure*}[!ht]
 \centering
	\footnotesize
	\setlength\tabcolsep{1pt} 
	\begin{tabular}{cccc}
	\textbf{(a) RF } & \textbf{(b) NN}  & \textbf{(c) SVR} & \textbf{(d) GPR} \\
 \IG[width=3 cm, width=3 cm, trim={0.0cm 0.0cm 0.0cm 0.0cm}, clip]{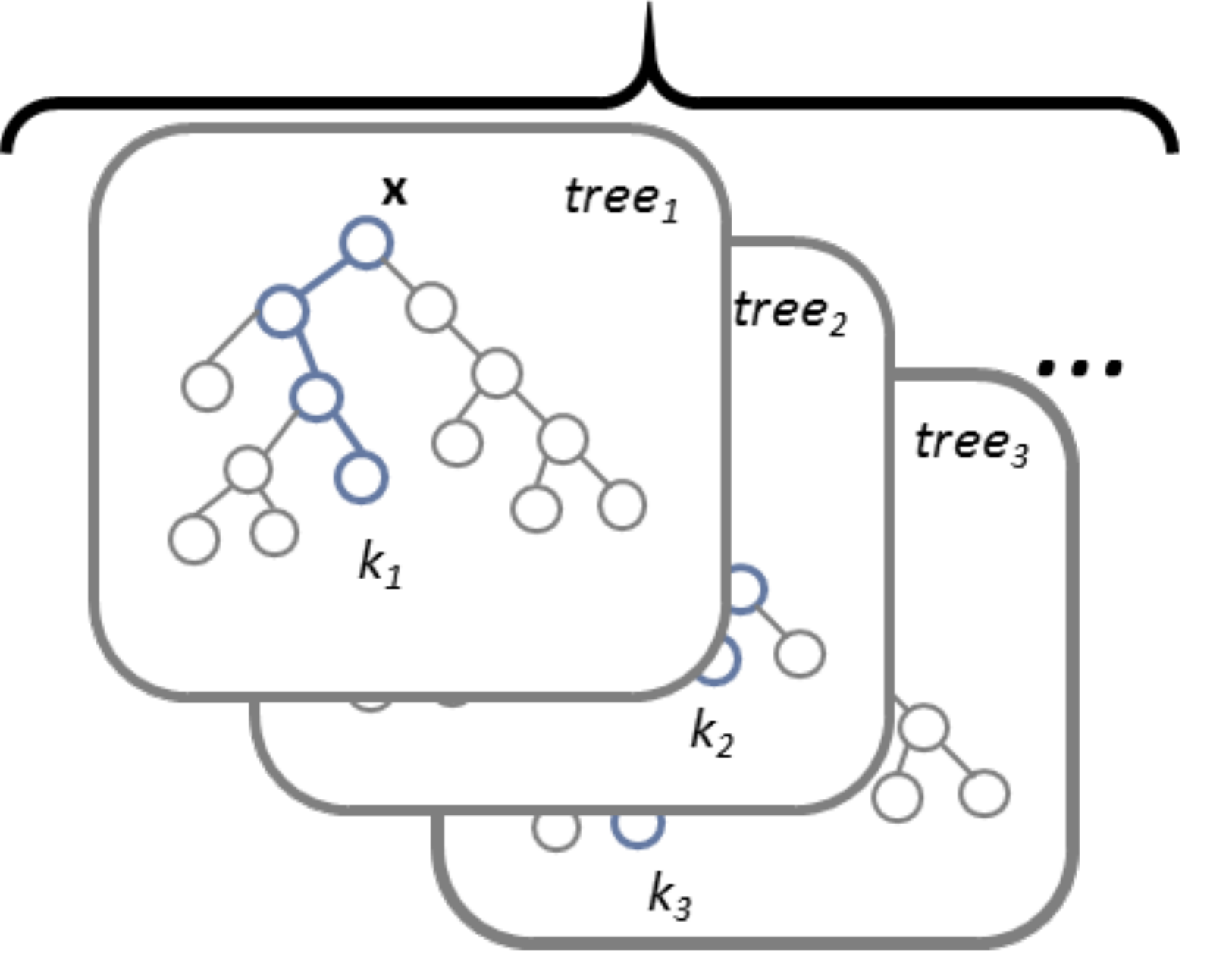} & 
 \IG[width=2.7 cm, width=3 cm, trim={0.0cm 0.0cm 0.0cm 0.0cm}, clip]{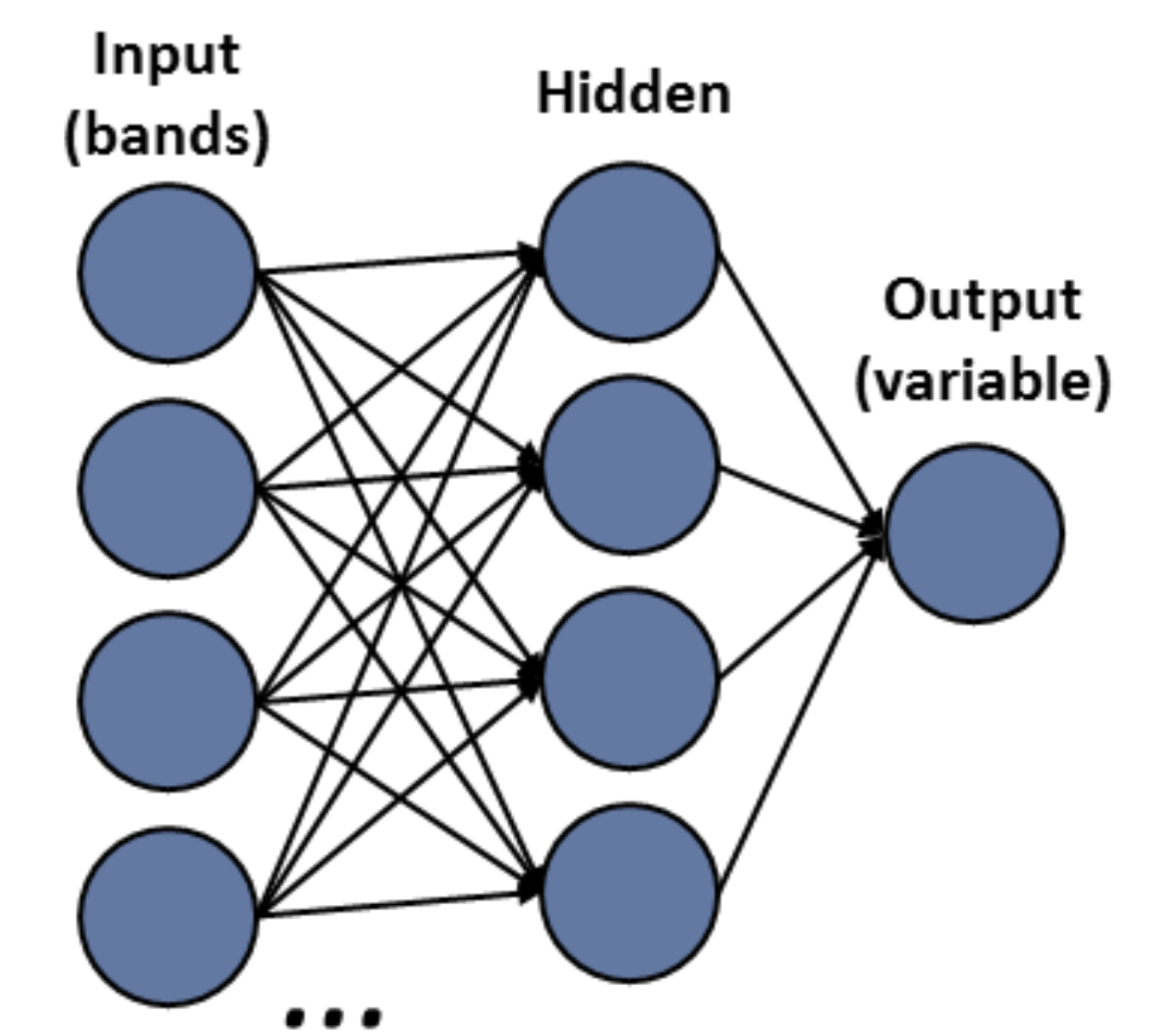} & 
 \IG[width=3.2 cm,width=3 cm, trim={0.0cm 0.0cm 0.0cm 0.0cm}, clip]{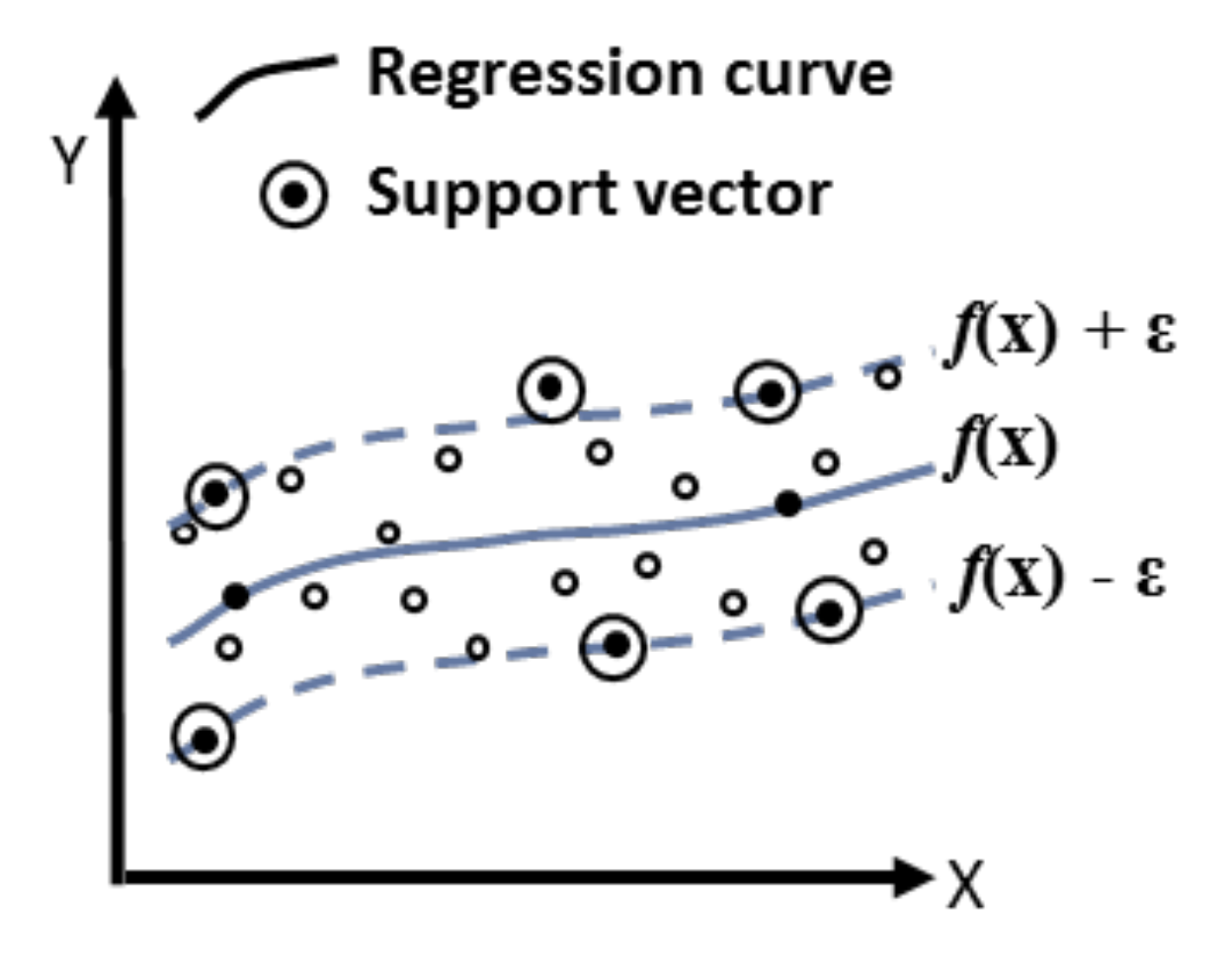} & 
 \IG[width=3 cm, width=3 cm, trim={0.0cm 0.0cm 0.0cm 0.0cm}, clip]{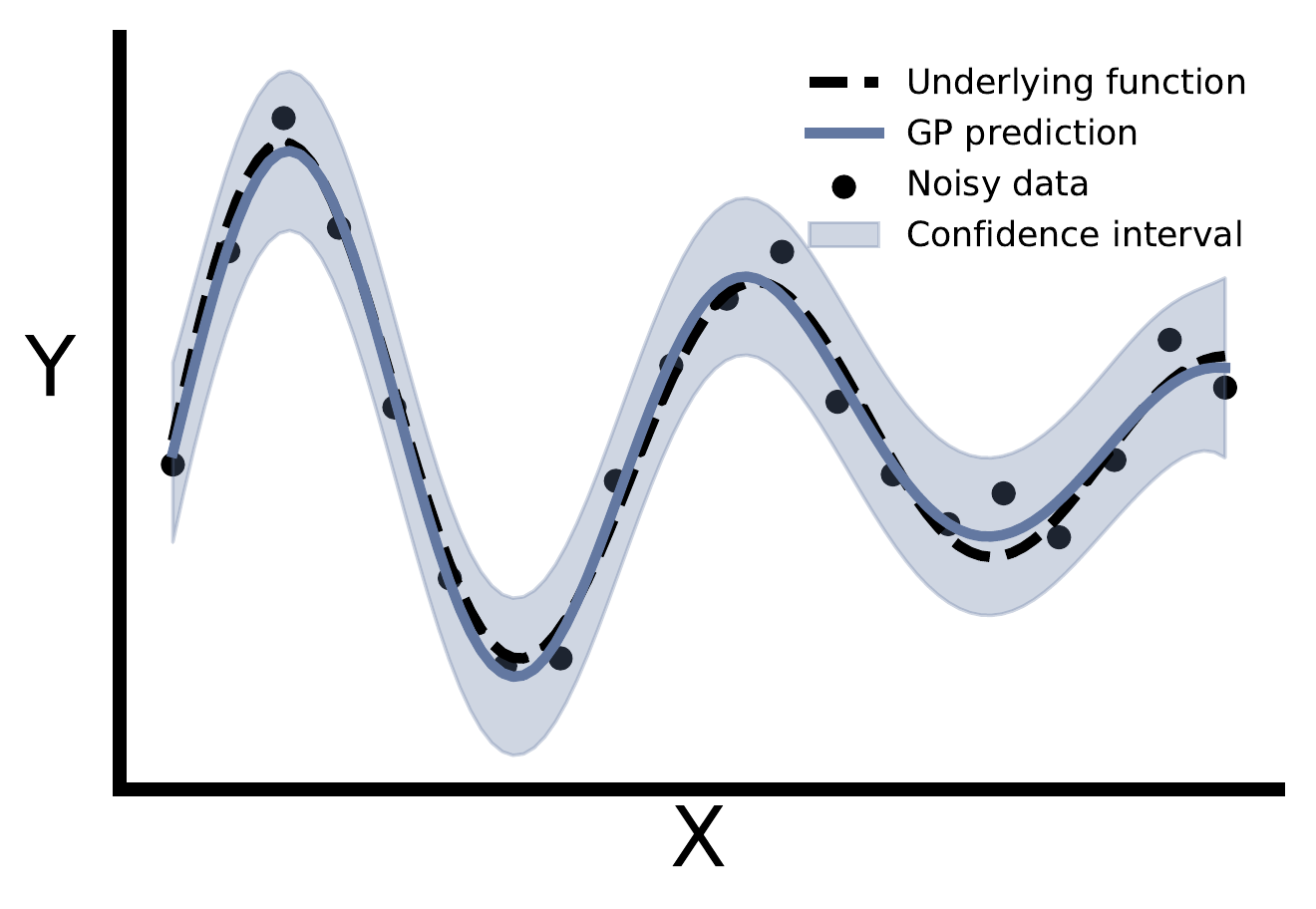} \\ 
    \end{tabular}\vspace{6pt}
     \caption{{Schematic illustrations of Random forest (RF) (a), Neural networks (NN) (b), Support vector regression (SVR) (c), and Gaussian processes regression (GPR) (d). }}
 \label{schematic_overview}
\vspace{-1.5em}
\end{figure*}

\section{Physically-based model inversion methods}  \label{sec:RTM}

Physically-based model inversion is based on physical laws establishing cause-effect relationships. Inferences on model variables are based on generally accepted knowledge embedded in radiative transfer models (RTMs). RTMs are deterministic models that describe absorption and multiple scattering, and some of them even describe the microwave region, thermal emission or sun-induced chlorophyll fluorescence emitted by vegetation (e.g., see Table \ref{advanced_RTMs}). A diversity of canopy RTMs have been developed over the last three decades with varying degrees of complexity. Gradual increase in RTMs accuracy, yet in complexity too, have diversified RTMs from simple turbid medium RTMs to advanced Monte Carlo RTMs that allow for explicit 3D representations of complex canopy architectures (e.g., see the RAMI exercises \citep{Pinty2001,Pinty2004,Widlowski2007,Widlowski2011,widlowski2015} for a thorough comparison). 
This evolution has resulted in an increase in the computational requirements to run the model, which bears implications towards practical applications. From a computational point of view, RTMs can be categorized as either (1) \emph{economically invertible} (or computationally cheap); or as (2) \emph{non-economically invertible} models (or computationally expensive). These terms refer to the model complexity and associated run-time constraining the mathematical inversion of such models. Economically invertible models are models with relatively few input parameters and fast processing that enables fast calculations and consequently fast model inversion or rendering of simulated scenes. A well-known example of this category includes the widely used leaf RTM PROSPECT \citep{Feret08} coupled with the canopy RTM SAIL \citep{Verhoef1984} (combined named as PROSAIL \citep{Jacquemoud2009}).

Non-economically invertible RTMs refer to advanced, computationally-expensive RTMs, often with a large number of input variables and sophisticated computational and mathematical modelling. These type of RTMs enable the generation of complex or detailed scenes, but at the expense of a significant computational load. In short, the following families of RTMs can be considered as non-economically invertible: (1) Monte Carlo ray tracing models (e.g., Raytran \citep{Govaerts1998}, FLIGHT \citep{North1996} and  librat \citep{Lewis1999}); (2) voxel-based models (e.g., DART \citep{gastellu1996}) and (3) advanced integrated vegetation and atmospheric transfer models (e.g., SCOPE \citep{VanDerTol2009} and MODTRAN \citep{Berk2006}). 
Descriptions of advanced canopy RTM models and their latest developments are provided in Table \ref{advanced_RTMs}.
Although these advanced RTMs serve perfectly as virtual laboratories for fundamental research on light-vegetation interactions, they are in general less suitable for retrieval applications, because of either a large number of input variables or a long processing time. Nevertheless, as outlined below, some experimental studies demonstrated that they can as well be applied into inversion schemes, e.g. based on look-up tables and in hybrid strategies. 

Regardless of their complexity, they all deliver spectroscopic outputs, typically at 1 nm resolution. Hence, RTMs outputs can fit perfectly into inversion schemes of imaging spectroscopy data, while at the same time the simulated data can be resampled to reassemble the band settings of multispectral sensors. Because inversion strategies are usually based on spectral fitting (i.e. only radiometric information is used), the drawback of collinearity complicating regression is not an issue here; however, removal of noisy bands is still a standard and much-needed preprocessing step to enable adequate spectral fitting. Another point to be mentioned is that inversion scheme can only retrieve the RTM input variables. Hence, using this strategy implies that only RTM state variables can be mapped. Yet because the RTM input variables drive the canopy absorbance and scattering mechanisms, the resulting output maps are considered to be physically sound \citep{Myneni1995a,knyazikhin2013}.  

\begin{figure*}[!h]
 \centering
	\footnotesize
	\begin{tabular}{c}
 \IG[width=12 cm, trim={0.0cm 1.0cm 0.0cm 0.0cm}, clip]{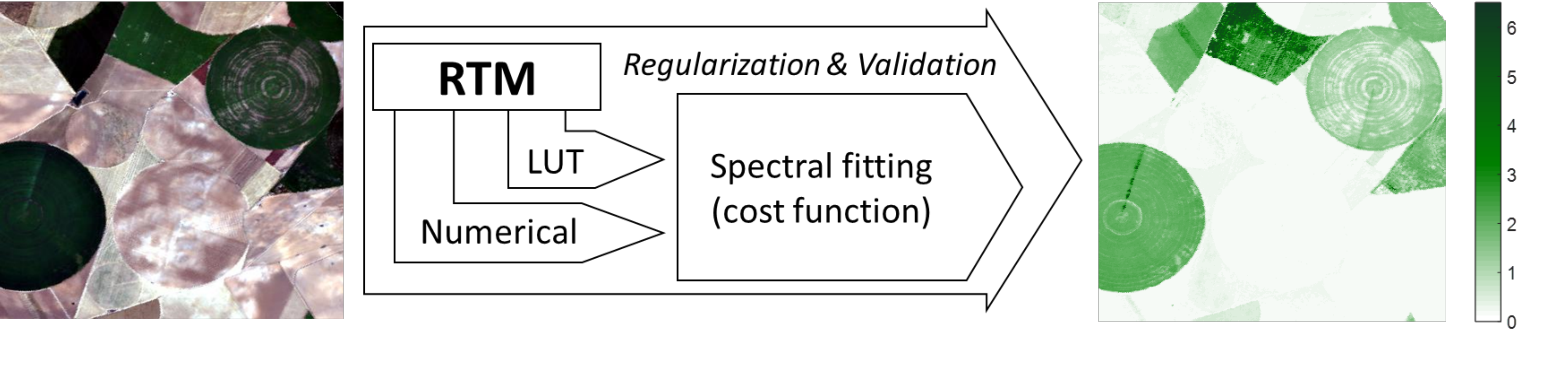} \\

    \end{tabular}\vspace{6pt}
     \caption{Principles of RTM inversion. Left: RGB subset of a hyperspectral HyMap image (125 bands) over Barrax agricultural site (Spain). Right: illustrative map of a vegetation property (LAI, m$^2$/m$^2$) as obtained by RMSE inversion against a 100000 PROSAIL LUT (5\% noise added, mean of 5\% multiple solutions). The model was validated with a R$^{2}$ of 0.44 (RMSE: 1.85; NRMSE: 31.9\%). A systematic underestimation occurred, which in principle implies that the RTM simulated LUT needs to be better parameterized.} It took 2315 seconds to produce the map using ARTMO's LUT-based inversion toolbox \citep{Rivera2013}. Also uncertainty estimates are provided, e.g. in the form of residuals (not shown). 
 \label{schematic_overview}
\vspace{-1.5em}
\end{figure*} 

\begin{table}[h]
\scriptsize
\centering
\caption{Advanded canopy RTMs commonly used in imaging spectroscopy applications.}
\label{advanced_RTMs}
\begin{tabular}{|>{\raggedright}p{1.3cm}|p{10cm}| }
\hline
\textbf{RTM} & \textbf{Description}  \\
\hline
SCOPE(Soil-Canopy-Observation of Photosynthesis and Energy fluxes) & 
SCOPE \citep{VanDerTol2009} is a Soil-Vegetation-Atmosphere (SVAT) scheme that includes RTMs along with a micrometeorological model for simulating turbulent heat exchange, and a plant physiological model for photosynthesis \citep{vandertol2014a}. The radiative transfer scheme is based on SAIL \citep{verhoef1984b, verhoef1985}, extended with a similar radiative transfer for emitted radiation. The emitted radiation includes chlorophyll fluorescence and thermal radiation. Leaf radiative transfer is calculated with Fluspect \citep{vilfan2016} which also includes emitted fluorescence radiation. SCOPE is intended as tool to scale processes from leaf to canopy, and to analyse the effects of light scattering. Recent developments include vertical heterogeneity \citep{yang2017} and the zeaxanthin-violaxanthin pigment cycles.  \\
\hline
Discrete Anisotropic Radiative Transfer (DART)  & DART model is being developed since 1992 as a physically based 3D computer programme \citep{gastellu1996}, which simulates radiative budget and remote sensing (airborne and space-borne) optical image data of natural and urban landscapes for any wavelengths from the ultraviolet to the thermal infrared part of the electromagnetic spectrum \citep{gastellu199, guillevic2003}. It computes and provides bottom and top of the atmosphere spectral quantities (i.e., irradiance, exitance and radiance) that are transformed into reflectance or brightness temperature depending on the user DART mode preferences \citep{gastellu2004}. 
Simulated scenes may include the atmosphere, topography and any natural or anthropogenic objects at any geographical location \citep{Grau2013}. The latest DART optical development includes also the specular reflectance and the light polarization  \citep{Gastellu2015}. Apart of passive remote sensing data, it also simulates active terrestrial and air-/space-borne light detection and ranging (LiDAR) discrete return, full waveform, multi-pulse and photon counting measurements  \citep{Gastellu2016,Yin2016}. 
In case of vegetation, it can also simulate radiative transfer of the solar-induced chlorophyll fluorescence for any virtual 3D Earth scene numerically and as images \citep{Gastellu-Etchegorry2017}. 
\\
\hline
librat & librat is a 3D Monte Carlo ray tracing radiative transfer model developed as a library interface to the original ararat (Advanced RAdiometric RAy Tracer) model. The first version of ararat was published in 1992 \citep{lewis1993} as part of the Botanical Plant Modelling System (BPMS) \citep{lewis1990,Lewis1999}. Subsequently, the sampling scheme was improved as reported in \citet{saich2002}, and the codes developed into a library in recent years. librat reads a 3D description of (canopy/soil/topographic) geometry, along with associated information on material scattering properties. The main function in the library then is that a ray is launched from some origin in 3D space, in a specified direction, and the code returns all information about the associated scattering paths and interactions, separated as direct and diffuse components. This core functionality, along with a set of associated sensor models but integrating paths fired into some volume. It allows for a wide range of radiative transfer calculations, including time-resolved/lidar, splitting of the radiometric information per scattering order etc. as well as straightforward raflectance/transmittance calculations \citep[e.g.][]{disney2006,hancock2012}. \\
\hline
FLIGHT & FLIGHT \citep{North1996,Barton2001} is a Monte Carlo ray tracing model designed to rapidly simulate light interaction with 3D vegetation canopies at high spectral resolution, and produce reflectance spectra for both forward simulation and for use in inversion \citep{Leonenko2013}. Foliage is represented by structural properties of leaf area, leaf angle distribution, crown dimensions and fractional cover, and the optical properties of leaves, branch, shoot and ground components. The model represents multiple scattering and absorption of light within the canopy and with the ground surface. It has been developed to model 3D canopy photosynthesis \citep{Alton2007}, to simulate waveform and photon counting lidar \citep{North2010,Montesano2015} and emitted fluorescence radiation \citep{Hernandez2017}. Structural data may be specified as a statistical distribution, derived from field measurements \citep{Morton2014} or by direct inversion from LiDAR data \citep{Bye2017}.\\
\hline
\end{tabular}
\end{table}

Given that in principle only a coupled leaf-canopy RTM and an inversion routine are required for the retrieval of RTM state variables, the approach is generic and generally applicable. Nevertheless, these approaches are not straightforward. First, an RTM has to be selected, whereby a trade-off between the realism and inversion possibility of the RTM has to be made. As discussed above, typically, complex models are more realistic, but they have many variables and consequently challenging to invert, whereas simpler models may be less realistic but easier to invert. Secondly, according to the Hadamard postulates, RTMs are invertible only when an inversion solution is unique and dependent -- in a continuous mode -- on the variables to be extracted. Unfortunately, this boundary condition is often not met. The inversion of canopy RTMs is frequently under-determined and ill-posed. The number of unknowns can be much larger than the number of independent observations. This makes physically-based retrievals of vegetation properties a challenging task. Several strategies have been proposed {to cope with}   
the under-determined problem of optimizing the inversion process, including (1) \emph{iterative numerical optimization} methods, (2) \emph{lookup-table (LUT) based inversion}, or (3) \emph{hybrid} approaches in which LUTs are generated as input for machine learning approaches (see section \ref{sec:hybrid}). Below we briefly review some common RTM inversion techniques in view of converting spectroscopic data into maps of RTM leaf and canopy input variables. 

\emph{Numerical optimization}: Iterative optimization is a classical technique to invert RTMs in image processing \citep{Jacquemoud95,Zarco01,Botha2007}. 
The optimization is minimizing a cost function, which estimates the difference between measured and estimated variables by successive input variable iteration. Optimization algorithms are computationally demanding and hence potentially time-consuming {depending on the complexity of the RTM and the numbers of image pixels to be processed.}
However, with the ongoing increase in computational power and open-source availability of optimization libraries, a renaissance of numerical approaches is emerging. Examples of numerical inversion against spectroscopic data include: PROSPECT inversion to retrieve leaf chlorophyll content \citep{Zhang2015}, retrieval of leaf biochemistry against an improved version of PROSPECT (COSINE) \citep{Jay2016}, and PROSAIL leaf and canopy variables \citep{Bayat2016, vandertol2016}. Despite a gain in computational power, numerical inversion algorithms applied to images are still time-consuming given the many per-pixel iterations and a high number of pixels involved. Hence, in its current form this method stays restricted to computationally fast RTMs in merely experimental settings. 

\emph{Look-up table (LUT)} strategies are based on the generation of simulated spectral reflectance scenarios for a high number of plausible combinations of variable value ranges.  
As such, the inversion problem is reduced to the identification of the modeled reflectance set that resembles most closely the measured one. This process is based on querying the LUT and applying a cost function. A cost function minimizes the summed differences between simulated and measured reflectances for all wavelengths. The main advantage of LUT-based inversion routines over numerical optimization is their computational speed, since the computationally most demanding part of the inversion procedure is completed before the inversion itself \citep{Dorigo2007}. {Consequently, LUT-based inversion methods are typically used as a preferred solution in RTM inversion studies.} The classical LUT-based inversion approach is based on a RMSE cost function, which continues to be applied until today. This approach proved to be especially successful for chlorophyll \citep{Zhang2008a,Kempeneers2008,Omari2013} and  LAI mapping. For instance, by using LUT-based inversion routines imaging spectroscopy data has been processed for the mapping of forest LAI \citep{Banskota2013, Banskota2015}, grassland LAI \citep{ Atzberger2015} and LAI over agricultural crops based on UAVs \citep{ Duan2014}. To further mitigate the ill-posed problem and optimize the robustness of the LUT-based inversion routines, {a diversity of } 
regularization strategies have been {explored} in inversion applications against spectroscopic data:

\begin{itemize}
  \item \emph{The use of prior knowledge} to constrain model variables in the development of a LUT \citep{Koetz05,Darvishzadeh08,Baret2008}. Prior knowledge typically involves information on the feasible variable ranges for involved vegetation types \citep{Dorigo09,Verrelst12c}. Prior information together with prior distributions are also increasingly applied into a Bayesian context, whereby the inverted values are generated based on likelihoods \citep{Laurent2013, Laurent2014,Shiklomanov2016}. The advantage of a Bayesian framework is its capability to quantify an inversion uncertainty around an inversion variable. 

  \item \emph{Selection of cost function}. The inverse problem of a nonlinear RTM is based on the minimization of a cost function concurrently measuring the discrepancy between (i) observed and simulated reflectance, and (ii) variables to estimate and the associated prior information \citep{Jacquemoud09}. To avoid solutions reaching fixed boundaries, a modified cost function in the LUT search that takes uncertainty of provided prior information into account is sometimes used, e.g. by means of the above-mentioned Bayesian approach. Alternatively, \citet{Leonenko2013} proposed and evaluated over 60 different cost functions dealing with different error distributions. Some more spectroscopic studies have evaluated among others the role of cost function \citep{Locherer2015, Danner2017} in LUT-based inversion. Although the classical RMSE is a robust cost function, sometimes improvements can be gained with alternative cost functions, e.g. when the LUTs are non-normal distributed.

  \item The use of \emph{multiple best solutions} in the inversion (mean or median), as opposed to a single best solution \citep{Koetz05,Locherer2015,Banskota2015,Kattenborn2017}. 

  \item The addition of \emph{artificial noise} (additive or inverse multiplicative white noise) to account for uncertainties linked to measurements and models \citep{Koetz05, Locherer2015,Danner2017}.
 
  \item Several spectroscopic studies reported that the relationship between measured and estimated variable perceptibly improves when only specific (sensitive) spectral ranges are selected for model inversion \citep{Meroni04,Schlerf05,Darvishzadeh12}. To account for noise in the observations, other spectroscopic studies instead manipulated the observed spectra by applying a smoothing filter \citep{Arellano2017} or  wavelet transforms \citep{Banskota2013,Ali2016,Kattenborn2017}. Spectral selection and spectral polishing methods can be applied at the same time in order to enhance the resemblance with the usually more spectrally smooth simulated spectra. 
  
\end{itemize}

Because of taking sun-target-sensor geometry into account, the use of RTM-based methods has been demonstrated to improve robustness to solar and view angle effects, compared to index-based methods \citep{Kempeneers2008}.  Another advantage of RTM inversion routines is that uncertainties are provided as spectral residuals \citep{Rivera2013} or standard deviations, when mapping multiple solution means \citep{Verrelst2014}.  
Yet the main drawback lies in its computational burden resulting from too many per-pixel iterations.  
Although LUT-inversion approaches may speed-up the inversion process as opposed to numerical inversion, these inversion routines are still computationally expensive due to the iterative calls of LUT entries on a per-pixel basis.  
 Consequently, despite attempts to optimize inversion algorithms in order to save-up computational time for solving inverse radiative transfer problems \citep{Gastellu2003,Favennec2016}, in terms of processing speed the RTM inversion routines run still behind statistical methods.

\vspace{-1.0em}
\begin{figure*}[!h]
 \centering
	\footnotesize
	\setlength\tabcolsep{3pt} 
	\begin{tabular}{cc}
	\textbf{(a) Numerical inversion} & \textbf{(b) LUT-based inversion} \\
 \IG[width=6.0 cm, trim={0.5cm 0.0cm 2.0cm 1.0cm}, clip]{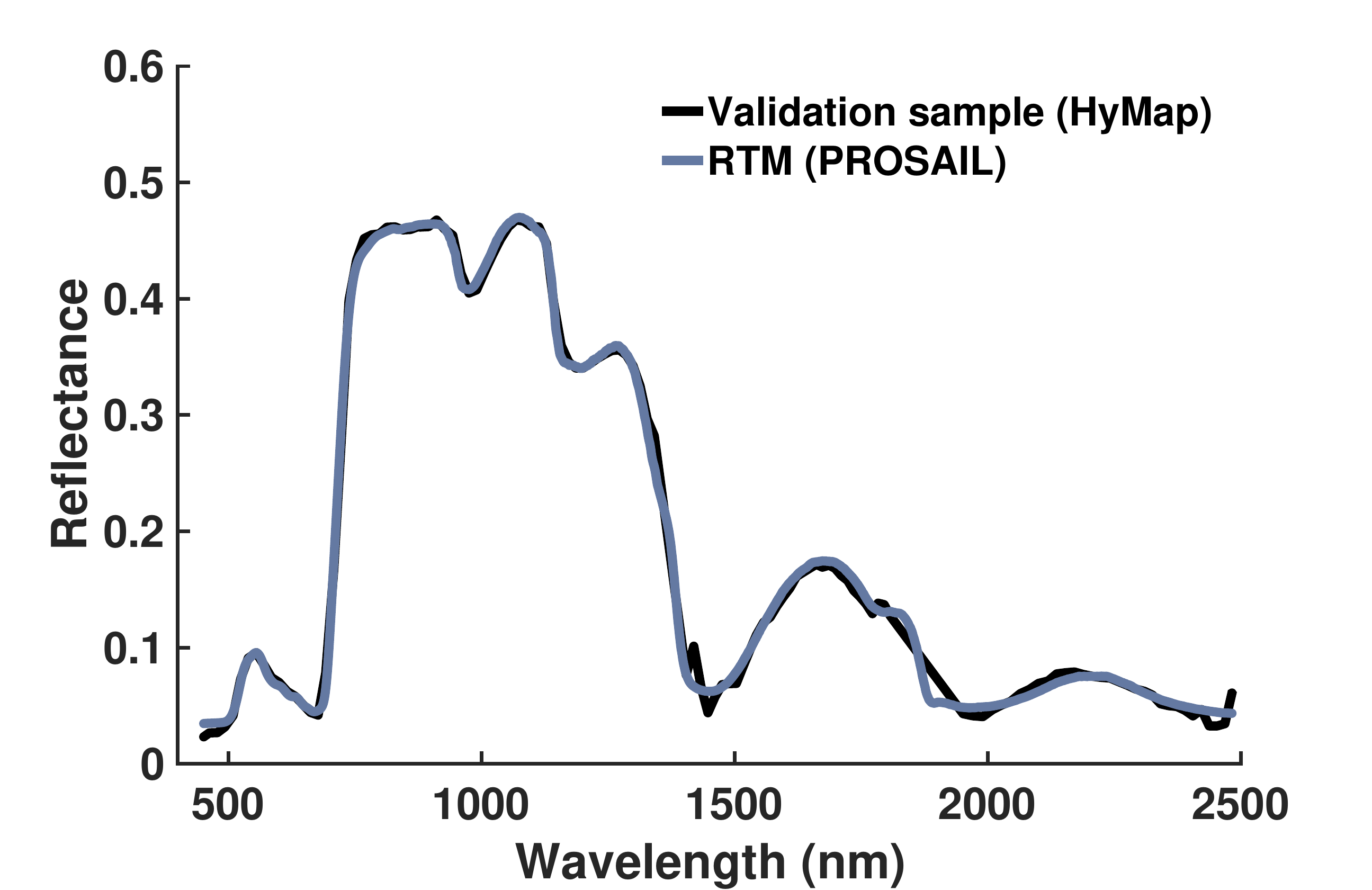} & 
 \IG[width=6.0 cm, trim={0.5cm 0.0cm 2.0cm 1.0cm}, clip]{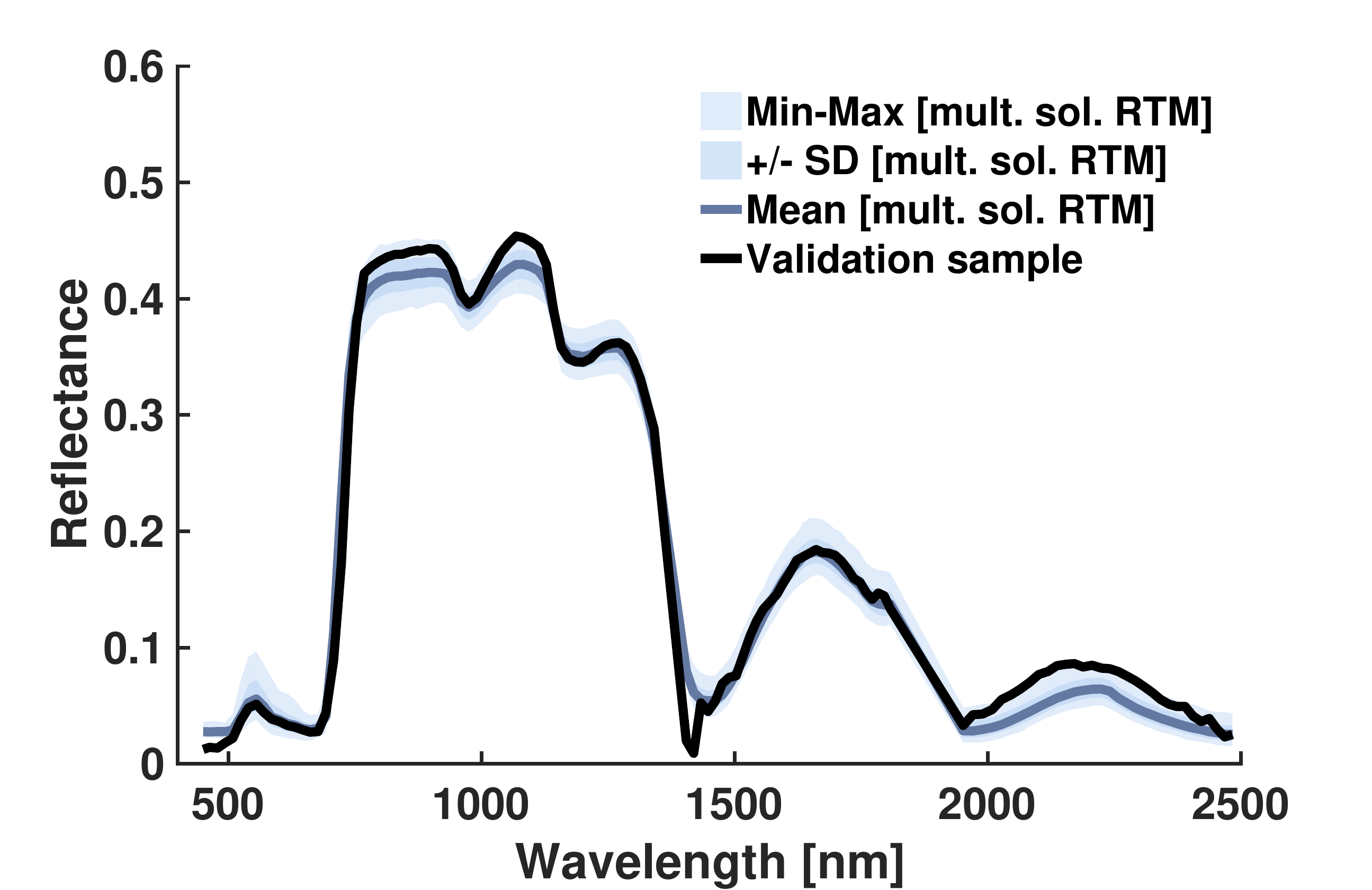} \\ 
    \end{tabular}\vspace{6pt}
     \caption{{Examples of numerical inversion (a) and LUT-based inversion (b). A HyMaP spectrum was inverted against PROSAIL. In case of LUT-inversion, overview statistics of 5\% best multiple solutions are shown.} }
 \label{schematic_overview}
\vspace{-1.5em}
\end{figure*}

\section{Hybrid regression methods}  \label{sec:hybrid}

Having discussed the more fundamental categories of retrieval methods, this section addresses \emph{hybrid} regression methods. Hybrid methods combine the generalization level of physically-based methods with the flexibility and computational efficiency of advanced machine learning methods. 
This approach replaces the ground data needed for training of the parametric or non-parametric models by RTM input variables, which makes it computationally efficient. It is important to note that the hybrid approach does not alleviate the main issues of RTMs, notably that they only include existing knowledge and concepts. 
Similarly as in case of LUT-based inversion, RTM simulations build a LUT representing a broad set of canopy realizations and the hybrid approach uses all available data stored in LUT to train a machine learning regression model.
  
\begin{figure*}[!h]
 \centering
	\footnotesize
		\begin{tabular}{c}
 \IG[width=12 cm, trim={0.0cm 1.0cm 0.0cm 0.0cm}, clip]{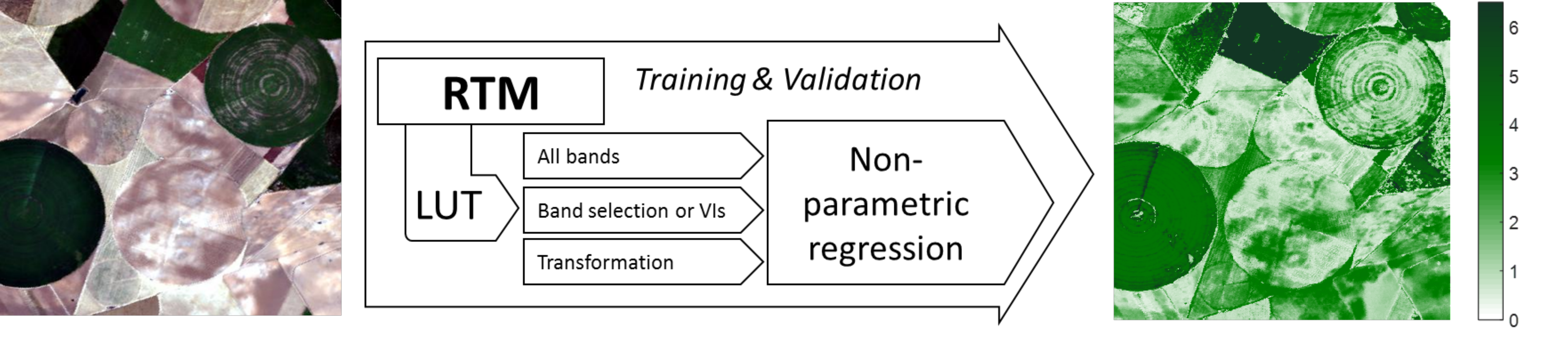} \\

    \end{tabular}\vspace{6pt}
     \caption{Principles of hybrid regression. Left: RGB subset of a hyperspectral HyMap image (125 bands) over Barrax agricultural site (Spain). Right: illustrative map of a vegetation property (LAI, m$^2$/m$^2$) as obtained by PROSAIL with Gaussian processes regression (GPR) and 15\% white noise added. The model was validated with a R$^{2}$ of 0.88 (RMSE: 0.70; NRMSE: 10.1\%)}. It took 6.3 seconds to produce the map using ARTMO's MLRA toolbox \citep{Caicedo2014}. With GPR also uncertainty estimates are provided (not shown). Because not being trained with bare soil spectra, LAI over the non-irrigated parcels is overestimated. 
 \label{schematic_overview}
\vspace{-1.5em}
\end{figure*}
 
The awareness in the mid 90's that ANNs are excellent algorithms to deal with large datasets  led to the introduction of hybrid methods based on ANNs trained with generically applicable RTM-generated data. It led to operational retrieval algorithms for datastreams acquired by multispectral and superspectral sensors (see \citet{verrelst2015b}). Although this approach is less straightforward in the context of imaging spectroscopy, because of the challenge of collinearity, some recent efforts have been undertaken in exploring this research direction. 
Noteworthy is the work of \citet{Vohland10} comparing a numerically optimized ANN with a LUT-based inversion using PROSAIL RTM simulations. Prediction accuracies generally decreased in the following sequence: numerical optimization > LUT > ANN. This would indicate that an ANN may not always be the best choice for inversion applications. However, no dimensionality reduction method was applied, which suggests that the regression model suffered from band collinearity effects. 
Also \citet{Fei2012} compared a PROSAIL-ANN hybrid approach with a PCA approach. The authors concluded that a PCA transformation into a regression function can mitigate the known reflectance saturation effect of dense canopies to some extent. 
This PROSAIL-ANN strategy was revisited by \citet{RiveraCaicedo2017} with alternative dimensionality reduction methods.   
Although PCA improved accuracies as opposed of using all bands, substantially more improvements were achieved when converting the spectra into components by means 
canonical correlation analysis (CCA) or orthonormalized PLS (OPLS).  

Likewise, inputs from more advanced RTMs were explored to develop specialized hybrid structures. In \citet{Malenovsky2013} an ANN was trained based on PROSPECT-DART simulations that explicitly took 3D canopy structures into account to estimate forest leaf chlorophyll content from hyperspectral airborne AISA data. In this approach the DART simulations went first through a continuum removal transformation.  
Alternatively, some studies have attempted to move away from ANN models by exploring hybrid structures on the basis of kernel-based machine learning regression algorithms, particularly the popular SVR. For instance, leaf chlorophyll content was estimated based on a PROSAIL-SVR model and applied to imaging spectroscopy \citep{Preidl2011}. An analogous concept was applied for a SVR that was trained by PROSPECT-DART simulations in combination with continuum removal transformations, with the purpose of quantifying forest biochemical and structural properties \citep{Homolova2016}. Similarly, \citet{Doktor2014} used a PROSAIL dataset to train a random forests (RF) model to predict LAI and leaf chlorophyll content, and \citet{Liang2016} compared PROSAIL-based hybrid models with SVR and RF for leaf and canopy chlorophyll content estimation from CHRIS data. Finally, \citep{RiveraCaicedo2017} analyzed ensembles of regression algorithms with dimensionality reduction methods to consolidate the most ideal PROSAIL-based (2101 bands) hybrid regression model. This study concluded that compressing PROSAIL data into CCA or OPLS components led to highest accuracies when trained with a GPR model. Altogether, although these studies have only been developed in experimental settings -- similar as the operational multispectral hybrid algorithms \citep[e.g.][]{Bacour06,Baret2013a} -- the hybrid structures can be perfectly implemented into global mapping schemes. 
When combined with a dimensionality reduction method to suppress collinearity, hybrid methods have a great potential to advance towards operational spectroscopy-based processing schemes. 

\section{Discussion}

The mapping of spatially continuous 
biophysical variables from imaging spectroscopy data is a progressively expanding field of research and development thanks to advances in spectrometer technology and in specialized methods interpreting the acquired spectral data. As a follow-up of an earlier, more general review on retrieval methods applicable to optical remote sensing \citep{verrelst2015b}, here a summary on retrieval methods specifically applied to spectroscopic data has been compiled. 
Four categories have been summarized: (1) parametric, (2) non-parametric, (3) RTM inversion, and (4) hybrid methods. The first two categories are statistical methods commonly used with experimental (field) data, whereas the latter two rely on RTM simulations. A schematic flowchart of the main retrieval methods and their hierarchy is provided in Figure \ref{schematic_overview}.
 
\begin{figure*}[!h]
 \centering
	\footnotesize
	\begin{tabular}{c}
 \IG[width=12 cm, trim={0.0cm 0.0cm 0.0cm 0.0cm}, clip]{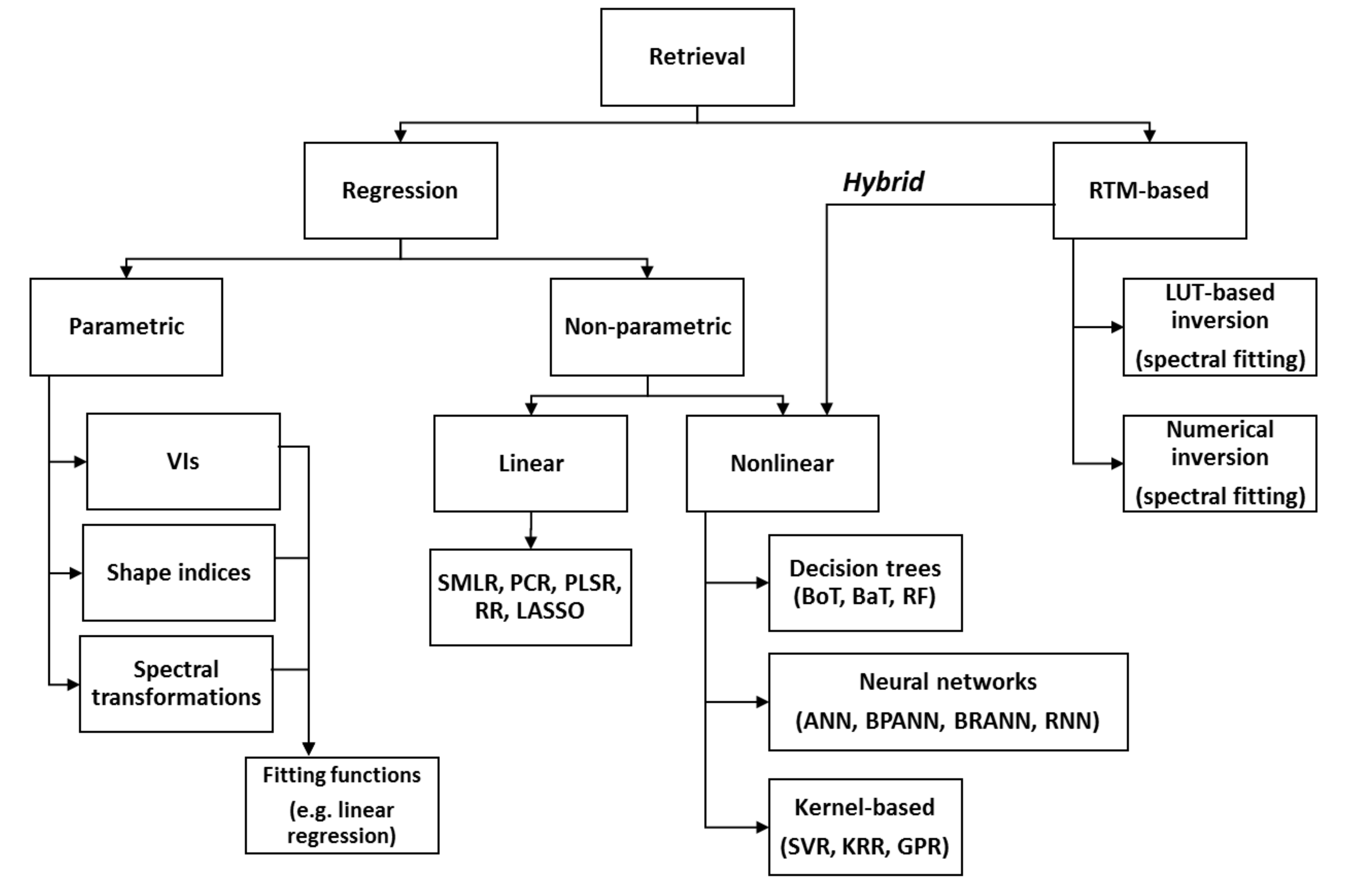} \\

    \end{tabular}\vspace{6pt}
     \caption{Schematic overview main retrieval methods.}
 \label{schematic_overview}
\vspace{-1.5em}
\end{figure*}

While pros and cons of each of these methodological categories have been earlier discussed \citep{verrelst2015b}, here we discuss these categories from the perspective of forthcoming routinely-acquired and standardized (e.g., atmospherically-corrected) imaging spectroscopy data streams.
First of all, the choice of a method bears implications; not only on the retrievability and processing time of mappable vegetation properties, but also on the purpose of the retrieval. 
Parametric and non-parametric methods rely on ground data for training, which obviously need to be available in order to apply these methods. If they are available, they are the 'shortest' way to the variables of interest, because especially the non-parametric methods do no impose any limitation on the relationship between the spectrum and the variable of interest. In contrast, RTMs describe radiative transfer processes, i.e. they use existing knowledge (as materialized in the models) rather than ground measured data. Retrieval from an RTM through inversion is most useful if one is more interested in the underlying radiative transfer processes (scattering, sun- and shade foliage fractions, light distribution within vegetation canopies, relationships between canopy structure and photosynthesis), rather than in merely extracting a specific variable. 
However, strategies relying on RTM simulations are inherently limited by the input variables of the RTM and, as discussed in section \ref{sec:RTM}, ancillary data and regularization methods may be required to optimize their inversions.  

Statistical approaches, on the other hand, possess the flexibility to relate reflectance data with any measured biophysical variable -- state variable or not. As has been demonstrated in sections \ref{sec:param} and \ref{sec:Nonparam}, this can be any quantifiable attribute, typically in the domains of leaf biochemical constituents (e.g., nitrogen, phosphorus), pigments (e.g., chlorophyll, cartenoids, xanthophylls) or higher-level structural variables (e.g., above-ground biomass, grain yield). 
The strength of the correlation with validation data typically determines the validity and transferrability of the statistical model. While this 'seeking for best correlations' can be criticized, because of the absence of a physical basis \citep{knyazikhin2013}, statistical approaches are becoming increasingly powerful to extract biochemical variables through complex and often indirect relationships. Particularly, machine learning models are powerful in extracting information from subtle variations in spectroscopic data through adaptive, nonlinear relationships. The advantage of these statistical models is that not only variable-specific absorption features can be used for information extraction, but also secondary relationships with variables related to other absorption features that co-vary with the variable of interest can be exploited \citep{Ollinger2011,Verrelst2012a}. Since high accuracies are often obtained with these methods, they are gaining popularity, not only for quantification of a diversity of vegetation properties, but also in mapping of floristic composition \citep{Roth2015,Harris2015,Neumann2016,Feilhauer2017}.

Regardless of the nature of retrieval method, in view of mapping larger areas, and especially in an operational and global context, what matters is the possibility to provide associated information on the retrieval quality. The characterization of uncertainty is a fundamental requirement for postulating correct scientific conclusions from results and for assimilating results into statistical or mechanistic higher-level models  \citep{Cressie2009}.
As addressed in section \ref{sec:param}, parametric regression methods, i.e. spectral transformation methods in combination with a fitting function, do not provide uncertainty estimates, which undermine their applicability to other images in space and time. Subsequently, while valid when locally calibrated and validated, parametric methods are of little use in an operational context. 
With regard to inversion routines, uncertainties can be provided as spectral residuals \citep{Rivera2013} or standard deviations when mapping multiple solution means \citep{Verrelst2014}. Lately, inversion approaches were proposed in a Bayesian framework \citep{Shiklomanov2016}, whereby uncertainties are delivered along with the retrievals. 
In case of traditional statistical models, uncertainty estimation has been a complex exercise. 
Statistical models developed within a Bayesian framework, such as GPR, provide uncertainties together with the predictions \citep{Verrelst2013d,CampsValls16grsm}, which indicate the probability interval of an estimation relative to the samples used during the training phase. These uncertainties can be used to evaluate GPR model transferability. For example, by mapping the uncertainties \citep{Verrelst2013d} demonstrated that a locally developed regression model can be successfully transported to other images in space and time for the large majority of pixels (i.e., the uncertainty maps were not systematically worse). Similarly, uncertainties can inform about the model performance. It was demonstrated that dimensionality reduction methods applied in GPR models for LAI mapping not only largely sped up the processing, but they also led to lower per-pixel uncertainties as opposed to mapping using all bands \citep{RiveraCaicedo2017}. {In conclusion, in the view of an operational processing need, just as important as the variable retrieval itself is the provision of an associated uncertainty estimate. Uncertainty estimates allow evaluating the method's per-pixel performance, and consequently allow evaluating the method's capability to process routinely acquired imaging data. They thus provide a measure of the retrieval fidelity, which can be used to identify and mask out the highly uncertain and non-reliable results. }

Another important aspect for operational production {of vegetation properties from typically bulky imaging spectroscopy data streams} implies computational speed. Generally, the lower the complexity of a model the faster it will be able to produce maps. This highly favors the application of parametric regression approaches since they consist of only few transformations and equations. Also non-parametric regression algorithms, once trained, can be applied to process an images almost instantaneously. Training of machine learning models is frequently related to the tuning of several free variables with costly cross-validation approaches. These scale poorly with the number of samples (such as in kernel machines) or with the data dimensions (such as in ANNs). Although a trained ANN converts an image into a map quasi-instantly, kernel-based methods require more processing time, because the similarity between each test pixel in the image and those used to train the model has to be estimated. Training can be computationally costly, especially when using a big  training dataset, e.g. as in hybrid strategies. A solution to  shorten training time could be in size reduction of the training data in a way that maximal relevant information is preserved. This can be achieved by means of dimensionality reduction methods in the spectral domain \citep{RiveraCaicedo2017}, or by means of intelligent sampling in the sampling domain, e.g. through active learning \citep{Verrelst2016AL}. 

Considerably longer run-time is expected in case of inversion routines. Since RTMs take some time to generate simulations, especially for computationally expensive models, and also the evaluation takes place on a per-pixel basis, the iterative inversion routines are computationally expensive leading to relatively slow mapping speeds. 
In an attempt to accelerate their mapping speed, it has been proposed to approximate the functioning of the original RTM by means of statistical learning called \emph{emulation} \citep{Gomez-Dans2015,Rivera2015}. Initial experiments to emulate leaf, canopy and atmospheric RTMs demonstrated that emulators can successfully generate spectral output from a limited set of input variable almost instantly, thereby preserving sufficient accuracy as compared to the original RTM \citep{Verrelst2016GSA,Verrelst2017}. Although an emulator reproduces RTM simulations instantly, application of a per-pixel spectral fitting requires many repetitions, which implies that these methods still do not reach the speed of statistical methods. 

All in all, having the purpose of advancing towards operational imaging spectroscopy data processing in mind, i.e., reaching  globally-applicable, accurate and fast estimates, we end up with the following recommendations:

\begin{itemize} [noitemsep,topsep=0pt]
  \item To enable model transferability to routinely-acquired images,  
  retrieval methods must provide associated per-pixel uncertainties as a quality indicator whether the model can perform adequately in another space and time.  

  \item  Regarding the computational speed, e.g. in case of repetitive image processing, statistical (i.e. regression) methods are multiple times faster than physically-based methods, capable to process full images in the order of minutes or even seconds. 

  \item In case of regression methods (experimental or hybrid), multicollinearity of spectroscopic data complicates the development of powerful models. Physically-based methods using spectral fitting do not suffer from this problem. 

  \item  To mitigate the problem of multicollinearity in regression methods, either band selection or dimensionality reduction methods can be applied before entering the regression. Although band selection is a common practice, likely more powerful regression models can be obtained when using a dimensionality reduction method. 
\end{itemize}

\section{Conclusions}\label{sec:Con}

With forthcoming imaging spectrometer satellite missions, an unprecedented stream of datasets on the terrestrial biosphere will become available. This will require powerful processing techniques enabling quantification of vegetation variables in an operational and global setting. Four categories of retrieval methods have been discussed in this review paper: (1) parametric regression; (2) non-parametric regression; (3) physically-based RTM inversion; and, (4) hybrid methods. For each of these categories, a diversity of methodological approaches are increasingly applied to imaging spectroscopy data.  
This literature review synthesized the current state-of-the-art in the field of spectroscopy-based vegetation properties mapping.  

Although parametric methods, such as shape indices or spectral transformation, deal well with extracting relevant information embedded in spectroscopic data, their lack of uncertainty estimates makes them unsuitable for operational use. 
Higher accuracies can be reached with nonlinear non-parametric methods; especially those in the field of machine learning that generate probabilistic outputs, e.g. Gaussian process regression. However, an additional step to mitigate their spectral multicollinearity is deemed necessary. 
A popular strategy in this respect is selecting a set of vegetation indices or applying spectral transformation before training the machine learning algorithm. It remains nevertheless questionable whether such band selection approaches fully capture all relevant information. Instead, dimensionality reduction methods that enable compressing the large majority of spectral variability into a few components tend to lead to more accurate predictions. 

On the other hand, the inversion of physically-based RTMs against spectroscopic data is generally applicable and physically sound, but optimizing their inversion strategies is more challenging compared to the regression methods. RTM-based inversion is computationally demanding and ancillary information is usually required as an input or to regulate the inversion algorithm.
Hybrid regression methods, based on the coupling of an RTM with a machine learning regression algorithm, overcome the problem of processing speed. Particularly Bayesian kernel-based hybrid strategies possess promising features, as they combine speed, flexibility and the provision of uncertainty estimates. Their accuracies and processing speed can be further improved in combination with dimensionality reduction. 
Altogether, and in the interest of operational spectroscopy-based mapping of vegetation properties, we recommend to further explore the feasibility and implementation of hybrid strategies into the next-generation data processing chains.

\end{document}